\documentclass[aps,prd,twocolumn,showpacs,nofootinbib]{revtex4}
\usepackage{epsfig,graphicx,amssymb,bm}
\newcommand{\simgt}{\lower.5ex\hbox{$\; \buildrel > \over \sim \;$}}
\newcommand{\simlt}{\lower.5ex\hbox{$\; \buildrel < \over \sim \;$}}

\newcommand{\colskip}{@{\hspace{0.3in}}}

\newcommand{\MNRAS}{Mon. Not. Roy. Astr. Soc.}

\begin{document}

\title{%
 Cosmology with High-redshift Galaxy Survey: Neutrino Mass and Inflation
}%
\author{%
 Masahiro Takada$^1$, Eiichiro Komatsu$^2$ and Toshifumi Futamase$^1$
}%
\address{%
 $^1$Astronomical Institute, Tohoku University, Sendai 980-8578, Japan
}%
\address{%
 $^2$Department of Astronomy, The University of Texas at Austin, 
Austin, TX 78712
}%


\begin{abstract}
 High-$z$ galaxy redshift surveys open up exciting possibilities
 for precision determinations of neutrino masses and inflationary
 models. The high-$z$ surveys are more useful for cosmology
 than low-$z$ ones
 owing to much weaker non-linearities in matter clustering, 
 redshift-space distortion and galaxy bias, which allows us to use the galaxy
 power spectrum down to the smaller spatial scales that are inaccessible 
 by low-$z$ surveys. We can then utilize the two-dimensional information of the
 linear power spectrum in angular and redshift space to measure
 the scale-dependent suppression of matter
 clustering due to neutrino free-streaming as well as the shape of the 
 primordial power spectrum. 
 To illustrate capabilities of high-$z$ surveys for constraining
 neutrino masses and the primordial power spectrum, we compare three future
 redshift surveys 
 covering 300~square degrees at $0.5<z<2$, $2<z<4$, and $3.5<z<6.5$.
 We find that, combined with the cosmic microwave background data
 expected from the Planck satellite, these surveys allow 
 precision determination of
 the total neutrino mass with the projected errors of  
 $\sigma(m_{\nu, {\rm tot}})=0.059$,
 0.043, and 0.025~eV, respectively, 
thus yielding a positive {\it detection} of the neutrino mass rather
 than an upper limit, as $\sigma(m_{\nu, {\rm tot}})$ is smaller than the 
 lower limits to the neutrino masses implied from the neutrino oscillation
 experiments, by up to a factor of 4 for the highest redshift survey.
 The accuracies of constraining the tilt and running index of the primordial
 power spectrum, 
 $\sigma(n_s)=(3.8,~3.7,~3.0)\times 10^{-3}$ 
 and  $\sigma(\alpha_s)=(5.9,~5.7,~2.4)\times 10^{-3}$ at $k_0=0.05~{\rm
 Mpc}^{-1}$, respectively, 
 are smaller than the current uncertainties
 by more than an order of magnitude, which will allow us to discriminate
 between candidate inflationary models.
 In particular, the error on  $\alpha_s$ from the future highest redshift survey
 is not very far away from the prediction
 of a class of simple 
 inflationary models driven by a massive scalar field
 with self-coupling, $\alpha_s=-(0.8-1.2)\times 10^{-3}$.
\end{abstract}
\maketitle

\section{Introduction}

We are living in the golden age of cosmology. 
Various data sets from precision measurements of temperature and
 polarization anisotropy in the
cosmic microwave background (CMB) radiation as well as 
those of matter density fluctuations in the large-scale structure
of the universe mapped by galaxy redshift surveys,
Lyman-$\alpha$ forests and weak gravitational lensing observations
are in a spectacular agreement 
with the concordance $\Lambda$CDM
model \cite{WMAP,Tegmark,Seljak,jarvis/jain:2005}. 
These results assure that theory of cosmological
linear perturbations is basically correct, and can accurately
describe the evolution of photons, neutrinos, baryons, and collisionless
dark matter particles \cite{Peebles,SZ,BE}, for 
given initial perturbations generated during 
inflation \cite{liddle/lyth:2000,dodelson}.
The predictions from linear perturbation theory can be compared with 
the precision cosmological measurements, in order to derive 
stringent constraints on the various basic cosmological parameters.  
Future observations with better sensitivity and higher precision will 
continue to further improve our understanding of the universe.

Fluctuations in different cosmic fluids 
(dark matter, photons, baryons, and neutrinos) 
imprint characteristic features in their power spectra, 
owing to their interaction properties, thermal history, equation of
state, and speed of sound.
A remarkable example is the acoustic oscillation in the photon-baryon
fluid that was generated 
before the decoupling epoch of photons, $z\simeq 1088$, which
has been observed in the power spectrum of CMB temperature 
anisotropy \cite{WMAPcl}, temperature--polarization cross 
correlation \cite{WMAPte},  and distribution of galaxies \cite{bao1,bao2}.

Yet, the latest observations have shown convincingly that we still do not 
understand much of the universe.
The standard model of cosmology tells us that the universe has been
dominated by four components. In chronological order the four
components are: early dark energy (also known as ``inflaton''
fields), radiation, dark matter, and late-time dark energy.
The striking fact is that we do not understand the precise nature
of three (dark matter, and early and late-time dark energy)
out of the four components; thus, understanding the nature of 
these three dark components has been and will continue to be one of the most important 
topics in cosmology in next decades.
Of which, one might be hopeful that the next generation particle
accelerators such as the Large Hadron Collider (coming on-line in 2007) 
would find some hints for the nature of dark matter particles.
On the other hand, the nature of late-time 
dark energy, which was
discovered by measurements of luminosity distance out to
distant Type Ia supernovae \cite{sn1,sn2}, is a complete mystery,
and many people have been trying to find a way to constrain 
properties of dark energy (see, e.g., \cite{DEreview} for a review).

How about the early dark energy, inflaton fields, which caused 
the expansion of the universe to accelerate in the very early universe?
We know little about the nature of inflaton, just like we know little
about the nature of late-time dark energy.
The required property of inflaton fields is basically the same as
that of the late-time dark energy component: both must have a large negative
pressure which is less than $-1/3$ of their energy density.
To proceed further, however, one needs more information from
observations.
Different inflation models make specific predictions for the shape of 
the power spectrum \cite{liddle/lyth:2000} (see also Appendix~B) 
as well as for other statistical
properties \cite{NGreview} of primordial perturbations.
Therefore, one of the most promising ways to constrain the physics
of inflation, hence the nature of early dark energy 
in the universe, is to determine the shape of the primordial power spectrum
accurately from observations. For example, the CMB data from 
the Wilkinson Microwave Anisotropy Probe \cite{WMAP}, 
combined with the large-scale
structure data from the Two-Degree Field Galaxy Redshift Survey \cite{2dF},
have already ruled out one of the popular inflationary models driven by
a self-interacting massless scalar field \cite{Peiris}.
Understanding the physics of inflation better will likely provide an
important implication for late-time dark energy.

``Radiation'' in the universe at around the matter-radiation equality 
mainly consists of photons and neutrinos; however, neutrinos 
actually stop being radiation when their mean energy per particle
roughly equals the temperature of the universe.
The physics of neutrinos has been revolutionized over the last decade 
by solar, atmospheric, reactor, and accelerator neutrino experiments having 
provided strong evidence for finite neutrino masses via
mixing between different neutrino flavors, the so-called neutrino 
oscillations \cite{exp1,exp2,exp3,exp4,exp5}. 
These experiments are, however, only sensitive to mass square 
differences between neutrino mass eigenstates, implying 
$\Delta m_{21}^2\simeq 7\times 10^{-5}$~eV$^2$ and
$\Delta m_{32}^2\simeq 3\times 10^{-3}$~eV$^2$; thus, the most fundamental
quantity of neutrinos, the absolute mass, has not been
determined yet.  
Cosmological
neutrinos that are the relic of the cosmic thermal history 
have distinct influences on the structure formation.
Their large energy density, comparable to the energy density of photons
before the matter-radiation equality, determines the expansion history 
of the universe.
Even after the matter-radiation equality, neutrinos having become
non-relativistic affect the structure formation by suppressing
the growth of matter density fluctuations at small spatial scales 
owing to their large velocity dispersion \cite{BES,BS,Pog1,Pog2,Ma95,Ma}
(see Sec.~II and Appendix~A for more details).
Therefore, the galaxy redshift surveys, combined with the CMB data, 
provide a powerful, albeit indirect, means to constraining 
the neutrino properties \cite{HET,Hannestad1,Lahav,LahavSuto,Hannestad2}.
This approach also complements the theoretical and direct experimental
efforts for understanding the neutrino physics.  In fact, the
cosmological constraints have placed the most stringent upper bound 
on the total neutrino mass, 
$m_{\nu,{\rm tot}}\simlt 0.6\,{\rm eV}$ ($2\sigma$) \cite{WMAPp}, 
stronger than the direct experiment limit $\simlt
2\,{\rm eV}$ \cite{Bonn}.
In addition, the result obtained from the 
Liquid Scintillator Neutrino Detector (LSND) experiment,
which implies $\bar{\nu}_\mu$ to $\bar{\nu}_e$
oscillations with $\Delta m^2 \simgt 0.2$\,eV$^2$ \cite{LSND}
in an apparent contradiction with the other neutrino oscillation
experiments mentioned above, potentially suggests the need for
new physics: the cosmological observations will provide independent
tests of this hypothesis.

In this paper we shall study the capability of future
galaxy surveys at high redshifts, combined with the CMB data, for 
 constraining (1) the neutrino properties, more specifically the 
total neutrino mass, $m_{\nu, {\rm tot}}$, and the number of 
non-relativistic 
neutrino species, $N_\nu^{\rm nr}$, and (2) the shape of the 
primordial power spectrum that is 
parameterized in terms of the spectral tilt,  $n_s$, and the running
index, $\alpha_s$, motivated by inflationary predictions (see
Appendix~B). 
For the former, we shall pay particular attention to our ability to
simultaneously constrain $m_{\nu, {\rm tot}}$ and $N_\nu^{\rm nr}$,
as they will provide important clues to resolving the absolute mass 
scale as well as the neutrino mass hierarchy.  
The accuracy of determining  the neutrino parameters and
the power spectrum shape parameters 
will be derived using
the Fisher information matrix formalism, including marginalization over 
the other cosmological parameters as well as the galaxy bias.

Our analysis differs from the previous work 
on the neutrino parameters in that 
we fully take into 
account the two-dimensional nature of the galaxy power spectrum in the
line-of-sight  and transverse directions, while the previous
work used only spherically averaged, one-dimensional power spectra. 
The geometrical distortion due to cosmology and the redshift space 
distortion due to the peculiar velocity field will cause anisotropic features 
in the galaxy power spectrum.
These features help to lift degeneracies between cosmological
parameters, substantially reducing the uncertainties in the 
parameter determinations.
This is especially true when variations in parameters of interest 
cause modifications in the power spectrum shape, which is indeed the 
case for the neutrino parameters, tilt and running index. 
The usefulness of the two-dimensional power spectrum, especially for 
high-redshift galaxy surveys, has been carefully investigated 
in the context of the prospected 
constraints on late-time dark energy properties
\cite{SE,Taka03,HuHaiman,Blake,GB05,taka05,Yamamoto}.

We shall show the parameter forecasts for future wide-field galaxy
surveys that are already being planned or seriously under consideration:
the Fiber Multiple Object Spectrograph (FMOS) on Subaru telescope
\cite{fmos}, its significantly expanded version, WFMOS \cite{wfmos}, the
Hobby--Ebery Telescope Dark Energy eXperiment (HETDEX) \cite{hetdex},
and the Cosmic Inflation Probe (CIP) mission \cite{cip}. 
To model these surveys, we consider three hypothetical galaxy surveys 
which probe the universe over different ranges of redshift, 
(1) $0.5\le z\le 2$, (2)
$2\le z\le 4$ and (3) $3.5\le z\le 6.5$. We fix the sky
coverage of each survey at $\Omega_{\rm s}=300$~deg$^2$ in order to 
make a fair comparison between different survey designs. As we
shall show below, high-redshift surveys are extremely powerful for 
precision cosmology because they allow
us to probe the linear power spectrum down to smaller length
scales than surveys at low redshifts, protecting the cosmological
information against systematics due to
non-linear perturbations.

We shall also study how the parameter uncertainties are affected by 
changes in the number density of sampled galaxies and the survey
volume. 
The results would  give us a good
guidance to defining the optimal survey design to achieve the desired
accuracies in parameter determinations. 

The structure of this paper is as follows. In Sec.~\ref{nu}, we review
the physical pictures as to how the non-relativistic (massive) neutrinos
lead to scale-dependent modifications in the growth of mass clustering
relative to the pure CDM model. 
Sec.~\ref{pps} defines the parameterization of the primordial power
spectrum motivated by inflationary predictions.
In Sec.~\ref{formalism} 
we describe a methodology to model
the galaxy power spectrum observable from
a redshift survey that includes the two-dimensional nature in the
line-of-sight and transverse directions.
We then present the Fisher
information matrix formalism that is used to estimate the projected 
uncertainties in the cosmological parameter determination from 
statistical errors on the galaxy power
spectrum measurement for a given survey. After survey parameters are
defined in Sec.~\ref{survey}, we show the parameter forecasts in 
Sec.~\ref{results}. Finally, we present conclusions and some discussions in 
Sec.~\ref{conc}.  We review the basic properties of cosmological
neutrinos in Appendix~A, the basic predictions from inflationary
models for the shape of the primordial power spectrum in Appendix~B,
and the relation between the primordial power spectrum and 
the observed power spectrum of matter density fluctuations in Appendix~C.

In the following, we assume an adiabatic, cold dark matter (CDM)
dominated cosmological model with flat geometry, which is 
supported by the WMAP results \cite{WMAP,WMAPp}, and employ the the 
notation used in \cite{HE98,HE99}: the
present-day density of CDM, baryons, and non-relativistic 
neutrinos, in units of the critical density, are denoted as 
$\Omega_{\rm c}$, $\Omega_{\rm
b}$, and $\Omega_{\nu}$, respectively. The total matter density is then
$\Omega_{\rm m}=\Omega_{\rm c}+\Omega_{\rm b}+\Omega_{\nu}$, and $f_\nu$
is the ratio of the massive neutrino density contribution to
$\Omega_{\rm m}$: $f_\nu=\Omega_\nu/\Omega_{\rm m}$.

\section{Neutrino Effect on Structure Formation}
\label{nu}

Throughout this paper we assume the standard thermal history in the
early universe: there are three 
neutrino species with temperature
equal to $(4/11)^{1/3}$ of the photon temperature.
We then assume that $0\leq N_\nu^{\rm nr}\leq 3$ species are massive and
could become
non-relativistic by the present epoch, and those non-relativistic
neutrinos have equal masses, $m_\nu$.
As we show in Appendix~\ref{app:neutrino}, the density parameter
of the non-relativistic neutrinos is given by
$\Omega_\nu h^2 = N^{\rm nr}_\nu m_\nu/(94.1\,{\rm eV})$, where
we have
assumed 2.725\,K for
the CMB temperature today \cite{mather},
and 
$h$ is the Hubble parameter defined as
$H_0=100\,h$\,km\,s$^{-1}$\,Mpc$^{-1}$. 
The neutrino mass fraction is thus given by
\begin{equation}
 f_\nu \equiv \frac{\Omega_\nu}{\Omega_{\rm m}}
= 0.05\left(\frac{N^{\rm nr}_\nu m_\nu}{0.658~{\rm eV}}\right)
\left(\frac{0.14}{\Omega_{\rm m}h^2}\right).
\label{eq:fnu}
\end{equation}

Structure formation 
is modified by non-relativistic neutrinos on scales 
below the Hubble horizon size when the neutrinos became non-relativistic, 
$k_{\rm nr}= 0.0145
(m_\nu/1\, {\rm eV})^{1/2}\Omega_{\rm m}^{1/2}h$\,Mpc$^{-1}$ 
(see Eq.~[\ref{eq:knr}]).
In particular, 
the characteristic scale imprinted onto the galaxy power spectrum at a
given redshift $z$ is the neutrino free-streaming scale, which is 
defined by Eq.~(\ref{eq:kfs}): 
\begin{equation}
 k_{\rm fs}(z)
=0.113~{\rm Mpc}^{-1} \left(\frac{m_\nu}{1~ {\rm eV}}\right)
\left(\frac{\Omega_{\rm m} h^2}{0.14}\frac5{1+z}\right)^{1/2}.
\label{eq:kfs*}
\end{equation}
Therefore, non-relativistic neutrinos with lighter masses suppress 
the growth of structure formation on larger spatial scales at a given redshift,
and the free-streaming length becomes shorter at a lower redshift as 
neutrino velocity decreases with redshift. 
The most important property of the free-streaming scale is that
it depends on the mass of each species, $m_\nu$, rather than the total
mass, $N_\nu^{\rm nr}m_\nu$; thus, measurements of $k_{\rm fs}$ allow us to distinguish
different neutrino mass hierarchy models. 
Fortunately, $k_{\rm fs}$ appears on the scales that are 
accessible by galaxy surveys: $k_{\rm fs}= 0.096-0.179~{\rm Mpc}^{-1}$
at $z=6-1$ for $m_\nu=1\,{\rm eV}$.  

On the spatial scales larger than the free-streaming length, 
$k< k_{\rm fs}$, neutrinos can cluster and fall into gravitational
potential well
together with CDM and baryonic matter.
In this case, perturbations in all matter components 
(CDM, baryon and neutrinos, denoted as `cb$\nu$' hereafter) 
grow at the same rate given by 
\begin{equation}
 D_{{\rm cb}\nu}(k,z)\propto D(z) \qquad k\ll k_{\rm fs}(z),
\end{equation}
where $D(z)$ is the usual linear growth factor 
(see, e.g., Eq.~(4) in \cite{TJ}).  
On the other hand, on the scales smaller than the free-streaming length,
$k> k_{\rm fs}$, perturbations in non-relativistic neutrinos are
absent due to the large velocity dispersion. In this case, 
the gravitational potential well is supported only by CDM and baryonic
matter, and the growth of matter perturbations is slowed
down relative to that on the larger scales.
As a result, the matter power spectrum for $k>k_{\rm fs}$ is 
suppressed relative to that for $k<k_{\rm fs}$. 
In this limit the total matter perturbations grow
at the slower rate given by
\begin{equation}
 D_{{\rm cb}\nu}(k,z) \propto (1-f_{\nu})[D(z)]^{1-p}\qquad k\gg k_{\rm fs}(z),
\label{eqn:D_cbnu}
\end{equation}
where $p\equiv (5-\sqrt{25-24f_\nu})/4$ \cite{BES}.  In \cite{HE98,HE99}
an accurate fitting function for the scale-dependent growth rate was
derived by matching these two asymptotic solutions.
We shall use the fitting function throughout this paper.

\begin{figure}
  \begin{center}
    \leavevmode\epsfxsize=8.cm \epsfbox{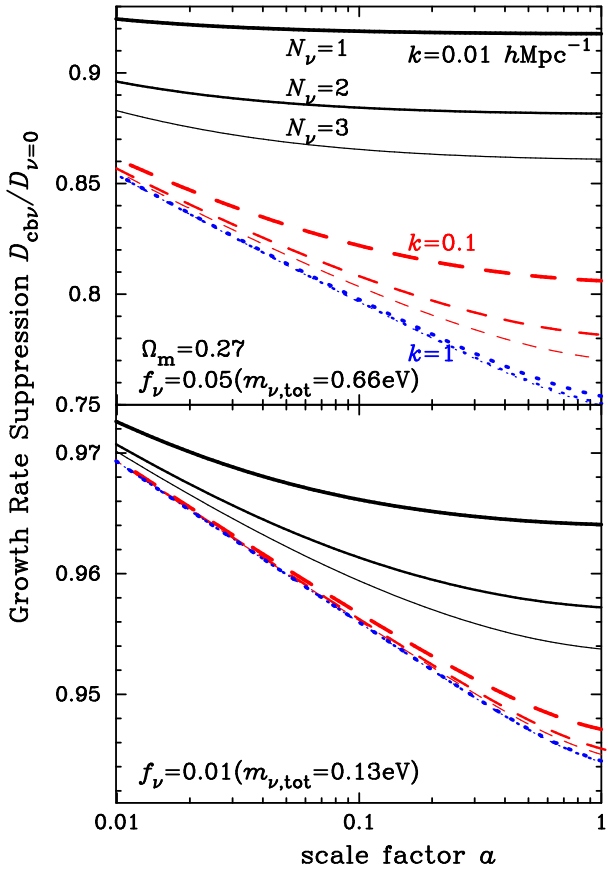}
  \end{center}
\vspace{-2em}
\caption{Suppression in the growth rate of total matter 
perturbations (CDM, baryons and
non-relativistic neutrinos), $D_{cb\nu}(a)$, 
due to neutrino free-streaming. ($a=(1+z)^{-1}$ is the scale factor.)
{\em Upper panel}: $D_{cb\nu}(a)/D_{\nu=0}(a)$ 
for the neutrino mass fraction of $f_\nu=\Omega_\nu/\Omega_{\rm m}=0.05$.
The number of non-relativistic neutrino
species is varied from $N_\nu^{\rm nr}=1$, 2, and 3
(from thick to thin lines), respectively.
The solid, dashed, and dotted lines represent $k=0.01$, 0.1, and 
1~$h$Mpc$^{-1}$, respectively.
{\em Lower panel}: $D_{cb\nu}(a)/D_{\nu=0}(a)$ 
for a smaller neutrino mass fraction,
$f_\nu=0.01$. Note that the total mass of non-relativistic
neutrinos is fixed to $m_{\nu,{\rm tot}}=N_\nu^{\rm nr}m_\nu=0.66$~eV
and 0.13~eV in the upper and lower panels, respectively.
}  \label{fig:growth}
\end{figure}

\begin{figure}
  \begin{center}
    \leavevmode\epsfxsize=8.5cm \epsfbox{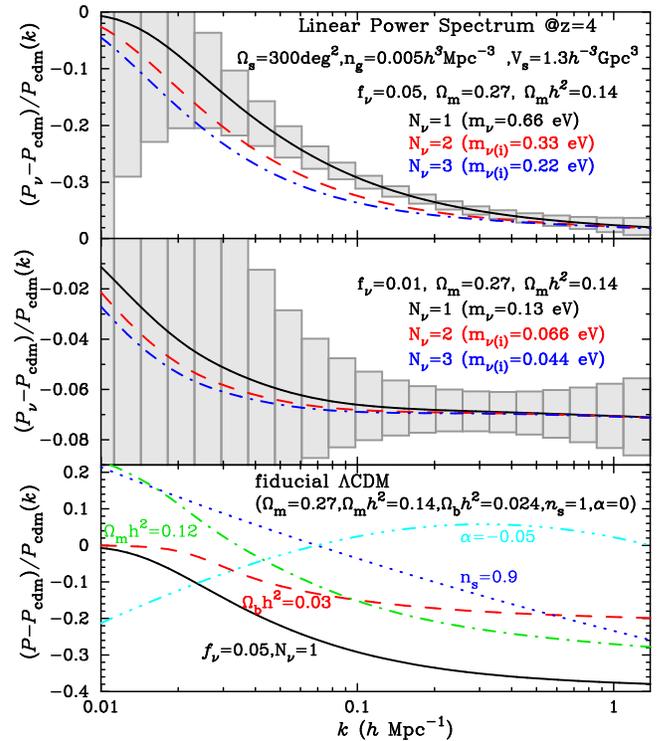}
  \end{center}
\vspace{-2em}
\caption{{\em Upper panel}: A fractional suppression of power 
 in the linear power spectrum at $z=4$ 
 due to free-streaming of non-relativistic neutrinos.
 We fix the total mass of non-relativistic neutrinos by 
 $f_{\rm \nu}=\Omega_\nu/\Omega_{\rm m}=0.05$, and vary 
 the number of non-relativistic neutrino species (which have equal
 masses, $m_\nu$) as $N^{\rm nr}_\nu=1$ (solid), 
 2 (dashed), and 3 (dot-dashed). The mass of individual neutrino
species therefore varies as $m_\nu=0.66$, 0.33, and 0.22~eV, respectively
 (see Eq.~[\ref{eq:fnu}]).
 The shaded regions represent the 1-$\sigma$
 measurement errors on $P(k)$ in each $k$-bin, expected from a galaxy
 redshift survey observing galaxies at $3.5\le z\le 4.5$
 (see Table \ref{tab:survey} for definition of the survey).  
Note that the errors are for the spherically averaged power spectrum
 over the shell of $k$ in each bin. 
Different $N_\nu^{\rm nr}$ could be discriminated in this case.
 {\em Middle
 panel}: Same as in the upper panel, but for a smaller neutrino
mass fraction, $f_\nu=0.01$. 
 While it is not possible to discriminate between different
$N_\nu^{\rm nr}$, the overall suppression on small scales is 
 clearly seen.
 {\em Lower panel}: Dependences of the shape of $P(k)$ on 
the other cosmological parameters.} \label{fig:pk}
\end{figure}

Figure \ref{fig:growth} shows suppression in the growth rate of 
total matter perturbations at $k=0.01$, 0.1, and 1~$h$Mpc$^{-1}$
due to the neutrino free-streaming.
The suppression becomes more significant at lower redshifts for a
given wavenumber, or for higher frequency perturbations at a given
redshift, 
because neutrino can grow together with CDM and baryonic
matter after the spatial scale of a given perturbation has become larger than the
neutrino free-streaming scale that varies with redshift as given by Eq.~(\ref{eq:kfs*}).
It is thus expected that a galaxy survey with different
redshift slices can be used to efficiently extract 
the neutrino parameters, $N^{\rm nr}_\nu$ and $m_\nu$.

The upper and middle panels of Figure \ref{fig:pk} illustrate how
free-streaming of non-relativistic neutrinos suppresses 
the amplitude of linear matter power spectrum, $P(k)$,
at $z=4$.
Note that we have normalized the primordial power spectrum
such that all the power spectra match at $k\rightarrow 0$ (see \S~\ref{pps}).  
To illuminate the dependence of $P(k)$ on $m_\nu$, 
we fix the total mass of non-relativistic neutrinos, $N_\nu^{\rm nr}m_\nu$, 
by $f_\nu=0.05$ and $0.01$ in the upper and middle panels,
respectively, and vary the number of 
non-relativistic neutrino species as $N^{\rm nr}_\nu=1$, $2$ and $3$.  
The suppression of power is clearly seen 
as one goes from $k<k_{\rm fs}(z)$ to 
$k>k_{\rm fs}(z)$ (see Eq.~[\ref{eq:kfs*}] for the value of $k_{\rm fs}$).
The way the power is suppressed may be easily understood
by the dependence of $k_{\rm fs}(z)$ on $m_\nu$; for example,
$P(k)$ at smaller $k$  is more suppressed 
for a smaller $m_\nu$, as lighter neutrinos have longer free-streaming lengths.
On very small scales, $k\gg k_{\rm fs}(z)$
($k\simgt 1$ and 0.1\,Mpc$^{-1}$ for $f_\nu=0.05$ and $0.01$,
respectively), however,
the amount of suppression becomes nearly independent of
$k$, and depends only on $f_\nu$ (or the total neutrino mass, $N^{\rm nr}_\nu m_\nu$) as
\begin{equation}
 \left|\frac{\Delta P}{P}\right| \approx 
2f_\nu\left[1+\frac{3\ln(D_{z=4})}5\right]
\approx 8f_\nu.
\label{eq:overallsuppression}
\end{equation}  
We therefore conclude that one can extract $f_\nu$ and $N^{\rm nr}_\nu$ 
separately from the shape of $P(k)$, if the suppression ``pattern''
in different regimes of $k$ is accurately measured from
observations.

Are observations good enough?
The shaded boxes in the upper and middle panels in Figure~\ref{fig:pk}
represent the 1-$\sigma$ measurement errors on $P(k)$ 
expected from one of the fiducial galaxy surveys outlined in
Sec.~\ref{survey}. 
We find that $P(k)$ will be measured with $\sim 1\%$
accuracy in each $k$ bin.  If
other cosmological parameters were perfectly known, the total mass
of non-relativistic neutrinos as small as
$m_{\nu,{\rm tot}}=N_\nu^{\rm nr}m_\nu
\simgt 0.001\,{\rm eV}$ would be detected at more
than 2-$\sigma$. This limit is much smaller than 
the lower mass limit
implied from the neutrino oscillation experiments, 0.06\,{\rm eV}.
This estimate is, of course, 
unrealistic because a combination of other cosmological 
parameters could mimic the
$N^{\rm nr}_\nu$ or $f_\nu$ dependence of $P(k)$.
The lower panel in Figure~\ref{fig:pk} illustrates
how other cosmological parameters change the shape of $P(k)$.
In the following,
we shall extensively study 
how well future high-redshift galaxy surveys, combined
with the cosmic microwave background data, can determine 
the mass of non-relativistic neutrinos and discriminate between
different $N_\nu^{\rm nr}$, 
fully taking into account degeneracies between cosmological
parameters.

\section{Shape of Primordial Power Spectrum and Inflationary Models}
\label{pps}

Inflation generally predicts that the primordial power spectrum of 
curvature perturbations is nearly scale-invariant.
Different inflationary models make specific predictions
for {\it deviations} of the primordial spectrum from a
scale-invariant spectrum, and the deviation is often parameterized
by the ``tilt'', $n_s$, and the ``running index'', $\alpha_s$, of 
the primordial power spectrum.
As the primordial power spectrum is nearly scale-invariant, 
$|n_s-1|$ and $|\alpha_s|$ are predicted to be much less than unity.

This, however, does not mean that the observed matter 
power spectrum is also nearly scale-invariant. 
In Appendix~\ref{app:norm}, we derive the power spectrum of
total matter 
perturbations that is normalized 
by the primordial curvature perturbation (see Eq.~[\ref{eq:pknorm}])
\begin{eqnarray}
 \frac{k^3P(k,z)}{2\pi^2}
&=&  \delta_{\cal R}^2\left(\frac{2k^2}{5H^2_0\Omega_{\rm m}}\right)^2
\nonumber \\
&& \hspace{-4em}\times
{D^2_{cb\nu}(k,z)}
T^2(k)\left(\frac{k}{k_0}\right)^{-1+n_s+\frac12\alpha_s\ln(k/k_0)},
\end{eqnarray}
where $k_0=0.05$~Mpc$^{-1}$,
$\delta^2_{\cal R}=2.95\times10^{-9}A$,
and $A$ is the normalization parameter given by
the WMAP collaboration \cite{WMAP}. We adopt
$A=0.871$, which gives $\delta_{\cal R}=5.07\times 10^{-5}$.
(In the notation of \cite{Hu02,HJ04} $\delta_{\cal R}=\delta_\zeta$.)
The linear transfer function, $T(k)$, describes the evolution 
of the matter power spectrum during radiation era and the interaction
between photons and baryons before the decoupling of photons.
Note that $T(k)$ depends only on non-inflationary parameters such 
as $\Omega_mh^2$ and $\Omega_b/\Omega_m$, and is independent of 
$n_s$ and $\alpha_s$. Also, the effects of non-relativistic neutrinos
are captured in $D_{cb\nu}(k,z)$; thus, $T(k)$ is independent of time
after the decoupling epoch.
We use the fitting function found in \cite{HE98,HE99} for $T(k)$.
Note that the transfer function and the growth rate
are normalized such that $T(k)\rightarrow 1$ 
and $D_{cb\nu}/a\rightarrow 1$ as $k\rightarrow 0$ 
during the matter era.

In Appendix~\ref{app:inflation} we
describe generic predictions on $n_s$ and $\alpha_s$ 
from inflationary models. 
For example, inflation driven by a massive, self-interacting scalar 
field predicts $n_s=0.94-0.96$ and $\alpha_s=(0.8-1.2)\times 10^{-3}$ for
the number of $e$-foldings of expansion factor before the end of inflation of 50.
This example shows that precision determination of $n_s$ and $\alpha_s$ 
allows us to discriminate between candidate inflationary models
(see \cite{liddle/lyth:2000} for more details).

\section{Modeling Galaxy Power Spectrum}
\label{formalism}

\subsection{Geometrical and Redshift-Space Distortion}
\label{distort}

Suppose now that we have a redshift survey of galaxies at some redshift.
Galaxies are biased tracers of the underlying gravitational field, and
the galaxy power spectrum measures how clustering strength of galaxies
varies as a function of 3-dimensional 
wavenumbers, $k$ (or the inverse of 3-dimensional length scales).

We do not measure the length scale directly in real space;
rather, we measure (1) angular positions of galaxies on the sky,
and (2) radial positions of galaxies in redshift space.
To convert (1) and (2) to positions in 3-dimensional space, 
however, one needs to assume a reference cosmological model, which
might be different from the true cosmology.
An incorrect mapping of observed angular and redshift 
positions to 3-dimensional positions
produces a distortion in the measured power spectrum,
known as the ``geometrical distortion'' \cite{AP,Taka96,Ballinger}.
The geometrical distortion can be described as follows.
The comoving size of an object at redshift $z$ in 
radial, $r_\parallel$, and transverse, $r_\perp$, directions are
computed from the extension in redshift, $\Delta z$, 
and the angular size, $\Delta \theta$, respectively, as
\begin{eqnarray}
r_\parallel&=& \frac{\Delta
 z}{H(z)},\nonumber\\
\label{eq:parallel}
r_\perp&=&D_A(z)\Delta \theta,
\label{eq:perp}
\end{eqnarray}
where $D_A$ is the comoving angular diameter distance
given in the spatial sector of the 
Friedmann-Robertson-Walker line element,
$dl^2=a^2(d\chi^2+D_A^2d\Omega)$ ($\chi$ is the comoving 
radial distance).
We assume a flat universe throughout this paper, 
in which case $\chi=D_A$.  The comoving angular distance out to a galaxy
at redshift $z$ is 
\begin{equation}
 D_A(z)=\int_0^z \frac{dz'}{H(z')},
\end{equation}
where $H(z)$ is the Hubble parameter given by
\begin{equation}
H^2(z)=H_0^2\left[\Omega_{\rm m}(1+z)^3+\Omega_{\Lambda}\right].
\end{equation}
Here $\Omega_{\rm m}+\Omega_{\Lambda}=1$, and $\Omega_\Lambda\equiv
\Lambda/(3H_0^2)$ is the present-day 
density parameter of a cosmological constant, $\Lambda$.
A tricky part is that $H(z)$ 
and $D_A(z)$ in Eq.~(\ref{eq:perp}) depend on cosmological
models. It is therefore necessary to assume some fiducial
cosmological model to compute the conversion factors.
In the following, quantities in the fiducial cosmological model
are distinguished by the subscript `fid'. 
Then, the length scales in Fourier space
in radial, $k_{{\rm fid} \parallel}$, 
and transverse, $k_{{\rm fid}\perp}$, directions are
estimated from the inverse of $r_{{\rm fid}\parallel}$ and 
$r_{{\rm fid} \perp}$. These fiducial wavenumbers are 
related to the true wavenumbers by
\begin{eqnarray}
k_\perp&=&\frac{D_A(z)_{\rm fid}}{D_A(z)}k_{{\rm fid}\perp},\nonumber\\
k_\parallel&=&\frac{H(z)}{H(z)_{\rm fid}}k_{{\rm fid}\parallel}.
\end{eqnarray}
Therefore, any difference between the fiducial cosmological model
and the true model would cause anisotropic distortions in the estimated
power spectrum in ($k_{\rm fid\perp}$, $k_{\rm fid\parallel}$)
space.

In addition, 
shifts in $z$ due to peculiar velocities of galaxies
distort the shape of the 
power spectrum along the line-of-sight direction, which
is known as the ``redshift space distortion'' \cite{Kaiser}. From
azimuthal symmetry around the line-of-sight direction, which is valid
when a distant-observer approximation holds, the linear
power spectrum estimated
in redshift space, $P_s(k_{\rm fid\perp}, k_{\rm fid\parallel})$, 
is modeled in \cite{SE} as
\begin{eqnarray}
P_s(k_{{\rm fid}\perp},k_{{\rm fid}\parallel})
&=&\frac{D_A(z)^2_{\rm fid}H(z)}{D_A(z)^2H(z)_{{\rm fid}}}
\left[1+\beta(k,z)\frac{k_\parallel^2}{k_\perp^2+k_\parallel^2}\right]^2\nonumber\\
&&\times b_1^2 P(k,z),
\label{eqn:ps}
\end{eqnarray}
where $k=(k_\perp^2+k_\parallel^2)^{1/2}$ and
\begin{equation}
 \beta(k,z)\equiv 
-\frac{1}{b_1}\frac{d\ln D_{{\rm cb}\nu}(k,z)}{d\ln (1+z)},
\end{equation}
is a function characterizing the linear redshift space distortion,
and $b_1$ is a scale-independent, linear bias parameter.  
Note that $\beta(k,z)$ depends on both redshift and wavenumber
via the linear growth rate.
In the infall regime, $k\ll k_{\rm fs}(z)$, we have 
$b_1\beta(k,z)\approx -d\ln D(z)/d\ln (1+z)$, while 
in the free-streaming regime, $k\gg k_{\rm fs}(z)$, we have
$b_1\beta(k,z) \approx -(1-p)d\ln D(z)/d\ln(1+z)$, where $p$ is defined
below Eq.~(\ref{eqn:D_cbnu}).

One might think that the geometrical and redshift-space distortion
effects are somewhat degenerate in the measured power spectrum. This 
would be true only if the power spectrum was a simple power law. 
Fortunately, characteristic, non-power-law features in $P(k)$
such as the broad peak from the matter-radiation equality, 
scale-dependent suppression of power due to baryons and non-relativistic
neutrinos, 
the tilt and running of the primordial power spectrum,
the baryonic acoustic oscillations, etc., help break 
degeneracies quite efficiently 
\cite{Taka96,Ballinger,SE,Taka03,HuHaiman,Blake,GB05,wfmos,taka05}.

\subsection{Comments on Baryonic Oscillations}
\label{sec:bao}

In this paper, we employ the linear transfer function 
with baryonic oscillations {\it smoothed out}
(but includes
non-relativistic neutrinos) \cite{HE98,HE99}. 
As extensively investigated in \cite{SE,wfmos,taka05}, 
the baryonic oscillations can be used as a standard
ruler, thereby allowing one to precisely constrain $H(z)$ and $D_A(z)$
separately 
through the geometrical distortion effects 
(especially for a high-redshift survey).
Therefore, our ignoring the baryonic oscillations might underestimate
the true capability of redshift surveys for constraining 
cosmological parameters.

We have found that the constraints on $n_s$ and $\alpha_s$
from galaxy surveys  improve by a factor of 2--3 when
baryonic oscillations are included.
This is because the baryonic oscillations basically fix 
the values of $\Omega_{\rm m}$, 
$\Omega_{\rm m}h^2$ and $\Omega_{\rm b}h^2$, lifting parameter
degeneracies between $\Omega_{\rm m}h^2$, $\Omega_{\rm b}h^2$, $n_s$, and $\alpha_s$.
However, we suspect that this is a rather optimistic forecast,
as we are assuming a flat universe dominated by a cosmological constant.
This might be a too strong prior, and 
relaxing our assumptions about geometry of the universe or the 
properties of dark energy will likely result in different forecasts
for $n_s$ and $\alpha_s$.
In this paper we try to separate the issues of non-flat universe
and/or equation of state of dark energy from the physics
of neutrinos and inflation.
We do not include the baryonic oscillations in our analysis,
in order to avoid too optimistic conclusions about the constraints
on the neutrino parameters, $n_s$, and $\alpha_s$.

Eventually, the full analysis including non-flat universe,
arbitrary dark energy equation of state and its time dependence,
non-relativistic neutrinos, $n_s$, and $\alpha_s$, 
using all the information we have at hand including the baryonic
oscillations, will be necessary. We leave it for a future publication
(Takada and Komatsu, in preparation).

\subsection{Parameter Forecast: Fisher Matrix Analysis}
\label{fisher}

In order to investigate how well one can constrain the cosmological
parameters for a given redshift survey design, one needs to 
specify measurement uncertainties of the galaxy power spectrum.
When non-linearity is weak, it is reasonable to assume that
observed density perturbations obey Gaussian statistics. In 
this case, there are two sources of statistical errors on a  
power spectrum measurement: the sampling variance (due to the
limited number of independent wavenumbers sampled from a finite survey
volume) and the shot noise (due to the imperfect sampling of fluctuations by
the finite number of galaxies). To be more specific, the
statistical error 
is given in \cite{FKP,Tegmark97} by
\begin{equation}
\left[\frac{\Delta P_s(k_i)}{P_s(k_i)}\right]^2= 
\frac{2}{N_k}\left[1+\frac1{\bar{n}_gP_s(k_i)}\right]^2,
\label{eqn:pkerror}
\end{equation}
where $\bar{n}_g$ is the mean number density of galaxies and 
$N_k$ is the number of independent $\bf{k}_{\rm fid}$ modes within
a given bin at $k_{\rm fid}=k_i$:
\begin{equation}
 N_k = 
2\pi k^2\Delta k\Delta\mu\left(\frac{2\pi}{V_s^{1/3}}\right)^{-3}.
\end{equation}
Here $2\pi/V_s^{1/3}$ is the size of the fundamental cell in
$k$ space, $V_s$ is the comoving survey volume, and 
$\mu$ is the cosine of the angle between $\bm{k}_{\rm fid}$ and the
line-of-sight.  
Note that we have assumed that the galaxy selection function 
is uniform over the redshift
slice we consider and ignored any boundary effects of survey geometry
for simplicity.  

The first term in Eq.~(\ref{eqn:pkerror}) represents sampling
variance. 
Errors become independent of the number density of galaxies
when sampling variance dominates (i.e., $P_s\gg \bar{n}_g$ over the
range of $k$ considered), 
and thus the only way to reduce
the errors is to survey a larger volume.
On the other hand, the second term represents shot noise, which 
comes from discreteness of galaxy samples. 
When shot noise dominates ($P_s\ll \bar{n}_g$), 
the most effective way to reduce noise
is to increase the number density of galaxies by increasing exposure time per field.
Note that for a fixed $\bar{n}_{g}$ the relative importance of 
shot noise contribution
can be suppressed by using galaxies with larger bias parameters, $b_1$,
as $P_s\propto b_1^2$.
In Sec.~\ref{survey} we shall discuss 
more about the survey design that is required to attain the desired parameter
accuracy.

We use the Fisher information matrix formalism to convert the errors on
$P_s(k)$ into error estimates of model parameters \cite{SE}. The Fisher
matrix is computed from
\begin{eqnarray}
F_{\alpha\beta}&=&\frac{V_{\rm s}}{8\pi^2}\int^{1}_{-1}\!\!d\mu 
\int^{k_{\rm max}}_{k_{\rm min}}k^2dk~ 
\frac{\partial \ln P_s(k,\mu)}{\partial
p_\alpha}
\frac{\partial \ln P_s(k,\mu)}{\partial p_\beta}\nonumber\\
&&\times 
 \left[\frac{\bar{n}_gP_s(k,\mu)}{\bar{n}_gP_s(k,\mu)+1}\right]^2,
\label{eqn:fisher}
\end{eqnarray}
where $p_\alpha$ expresses a set of parameters.
One may evaluate some derivative terms analytically:
\begin{eqnarray}
 \frac{\partial \ln P_s(k,\mu)}{\partial \delta_{\cal R}}
&=&
 \frac2{\delta_{\cal R}},\\
 \frac{\partial \ln P_s(k,\mu)}{\partial n_s}
&=& \ln\frac{k}{k_0},\\
 \frac{\partial \ln P_s(k,\mu)}{\partial \alpha_s}
&=& \frac12\left(\ln\frac{k}{k_0}\right)^2.
\end{eqnarray}
The $1\sigma$ error on $p_\alpha$ marginalized over the other parameters
is given by $\sigma^2(p_\alpha)=(\bm{F}^{-1})_{\alpha\alpha}$,
where $\bm{F}^{-1}$ is the inverse of the Fisher matrix.  It is sometimes
useful to consider projected constraints in a two-parameter subspace to
see how two parameters are correlated. 
We follow the method described 
around Eq. (37) in \cite{TJ} for doing this.  
Another
quantity to describe degeneracies between given two parameters, 
$p_\mu$ and $p_\nu$,
is the correlation coefficient defined as
\begin{eqnarray}
r(p_\alpha,p_\beta)=\frac{({\bm F}^{-1})_{\alpha\beta}}
{\sqrt{({\bm{F}}^{-1})_{\alpha\alpha}({\bm{F}}^{-1})_{\beta\beta}}}.
\label{eqn:coeff}
\end{eqnarray}
If $|r|=1$, the parameters  are totally degenerate, while
$r=0$ means they are uncorrelated. 

To calculate $F_{\alpha \beta}$ using Eq. (\ref{eqn:fisher}), we need to specify
$k_{\rm min}$ and $k_{\rm max}$ for a given galaxy survey. We use
the upper limit, $k_{\rm max}$, to exclude information in the non-linear
regime, where the linear theory prediction of density fluctuations,
Eq.~(\ref{eqn:ps}), becomes invalid.
Following
\cite{SE}, we adopt a conservative estimate for $k_{\rm max}$ by
imposing the condition $\sigma_{\rm mass}(R,z)=0.5$, 
where $\sigma_{\rm mass}(R,z)$ is the r.m.s.
mass fluctuation in a sphere of radius $R=\pi/(2k_{\rm max})$ at a given
redshift $z$. All the Fourier modes below $k_{\rm max}$ are considered as in
the linear regime. This idea is partly supported by the simulation-based work in the
literature \cite{Meiksin,white05,SE05}, while a more careful and quantitative
study is needed to understand the impact of non-linearities on cosmological
parameter estimates as well as to
study how to protect the cosmological information
against the systematics. 
Table \ref{tab:survey} lists $k_{\rm max}$ for
each redshift slice of galaxy surveys we shall consider.  
In addition, we shall show how the results will change with 
varying $k_{\rm max}$.  
As for the minimum wavenumber, we use 
$k_{\rm min}=10^{-4}$ Mpc$^{-1}$, which gives 
well-converged results for all the cases we consider.

\subsection{Model Parameters}
\label{model}

The parameter forecasts derived from the Fisher information formalism
depend on the fiducial model and are also sensitive to the choice of 
free parameters. We include a fairly broad range of the CDM dominated
cosmology:
the density parameters are $\Omega_{\rm m}(=0.27)$, $\Omega_{\rm m}h^2(=0.14)$,
and $\Omega_{\rm b}h^2(=0.024)$ (note that we assume a flat universe);
the primordial power spectrum shape parameters are
the spectral tilt, $n_s(=1)$, the running index, $\alpha_s(=0)$, 
and the normalization of primordial curvature perturbation, 
$\delta_{\cal R}(=5.07\times 10^{-5})$  
(the numbers in the parentheses denote the values of the fiducial model). 
The linear bias parameters, 
$b_1$, are calculated for each redshift
slice as given in Table~\ref{tab:survey}; 
the fiducial values of the neutrino parameters, $f_\nu$ and $N^{\rm nr}_\nu$,
are allowed to vary in order to study how the constraints on $f_\nu$ and
$N^{\rm nr}_\nu$ change with the assumed fiducial values.  
For a survey which consists of $N_s$ redshift slices, 
we have $8+N_s$ parameters in total.  

As we shall show later, a galaxy survey alone cannot determine all
the cosmological parameters simultaneously, but would leave some
parameter combinations degenerated. This is especially true when 
non-relativistic neutrinos are added.
Therefore, it is desirable to combine the galaxy survey constraints 
with the constraints from CMB temperature and polarization anisotropy,
in order to lift parameter degeneracies.
When computing the
Fisher matrix of CMB, we employ 7 parameters: 6 parameters (the
parameters above minus the neutrino parameters and the bias parameters) 
plus the Thomson scattering
optical depth to the last scattering surface, $\tau(=0.16)$.  
Note that we ignore the effects of non-relativistic neutrinos
on the CMB power spectra: their effects are small and do not add
very much to the constraints from the high-$z$ galaxy survey.
We then add the CMB Fisher matrix to the galaxy Fisher matrix as
$F_{\alpha\beta}=F^{\rm g}_{\alpha\beta} +F^{\rm CMB}_{\alpha\beta}$. 
We entirely ignore the contribution to the CMB from
the primordial gravitational waves.
We use the publicly-available CMBFAST code
\cite{cmbfast} to compute the angular power spectrum of
temperature anisotropy, $C^{\rm TT}_l$,
$E$-mode polarization,  $C^{\rm EE}_l$, and their cross 
correlation, $C^{\rm TE}_l$.
Specifically we consider the noise per pixel and the angular resolution
of the Planck experiment that were assumed in \cite{EHT}.  Note that we
use the CMB information in the range of multipole $10\le l\le 2000$.

\section{Galaxy Survey Parameters}
\label{survey}

\begin{table*}
\begin{center}
\begin{tabular}{lcccccccc}
\hline\hline
&  & $k_{\rm max} $ & $\Omega_{\rm survey}$ & $V_{\rm s}$ &
 $\bar{n}_{\rm g}$ & 
\hspace{1em}Bias\hspace{1em} 
 &$P_g\bar{n}_g$
\\
Survey & $z_{\rm center}$ &($h$Mpc$^{-1}$) & (deg$^2$) &
($h^{-3}$Gpc$^3$) 
& ($10^{-3}~ h^3$Mpc$^{-3}$)& 
&($k_{\rm max}$)
\\
\hline 
G1 ($0.5<z<2$) & $0.75$ & $0.14 $& 300 & 0.33 & 0.5 & 1.22 & 4.83\\
    & $1.25$ & $0.19$ & 300 & 0.53 & 0.5 & 1.47&2.49\\
    & $1.75$ & $0.25$ & 300 & 0.64 & 0.5 & 1.75&1.38\\
G2 ($2<z<4$) & $2.25$ & $0.32$& 300 & 0.68 & 0.5 & 2.03&0.80\\
    & $2.75$ &$0.41$& 300 & 0.69 & 0.5 & 2.32&0.46\\
    & $3.25$ &$0.52$& 300 & 0.67 & 0.5 & 2.62&0.27\\
    & $3.75$ &$0.64$& 300 & 0.64 & 0.5 & 2.92&0.16 \\
SG ($3.5<z<6.5$)  & $4$ &$0.71$& 300 & 1.26 & 5 & 4&2.19 \\
  & $5$ &$1.01$& 300& 1.13 & 5 &5 &1.04\\ 
  & $6$ &$1.50$& 300& 1.02 & 5 &5.5 &0.35\\
\hline
\end{tabular}
\end{center}
\caption{\label{tab:survey} Galaxy survey specifications that we assume in
this paper (see Sec.~\ref{survey} for the details).  We assume a fixed sky
coverage (300 deg$^2$) for all the surveys, and $V_{\rm s}$ and
$\bar{n}_{\rm g}$ are the comoving survey volume and the comoving number density
of sampled galaxies for each redshift slice, respectively.  $z_{\rm center}$ denotes the
center redshift of each redshift slice, and $k_{\rm max}$ is the maximum
wavenumber below which information in the linear power spectrum can
be extracted. (We do not use any information above $k_{\rm max}$ in
the Fisher information matrix analysis.) ``Bias'' denotes the 
assumed linear bias parameters of sampled galaxies.}
\end{table*}
We define the parameters of our hypothetical galaxy surveys 
so that the parameters resemble the future surveys that are 
already being planned and seriously pursued.  As shown in
Eq.~(\ref{eqn:pkerror}), the statistical error of the galaxy power
spectrum measurement is limited by the survey volume, $V_{\rm s}$, 
as well as the mean number density of galaxies, $\bar{n}_g$.  
There are two advantages for the high-redshift galaxy surveys.  
First, given a fixed solid
angle, the comoving volume in which we can observe galaxies is larger at
higher redshifts than in the local universe. Accordingly, it would be
relatively easy to obtain the well-behaved survey geometry, e.g.,
a cubic geometry that would be helpful to handle the systematics.
Second, density fluctuations at smaller spatial scales are still in the linear
regime or only in the weakly non-linear regime at higher
redshift, which gives us more leverages on measuring the shape of the 
linear power spectrum.

Of course, we do not always win by going to higher redshifts.
Detecting galaxies at higher redshifts is obviously more observationally
demanding,
as deeper imaging capabilities and better sensitivity for
spectrographs  are required. 
To increase the survey efficiency, the use of Multi Object Spectrographs (MOS) or
Integral Field Unit (IFU) spectrographs will be favorable. 
It is therefore unavoidable to have a trade-off in the survey design
between the number of spectroscopic targets and the survey volume: 
for a fixed duration of the survey (or a fixed amount of budget),
the total number of spectroscopic targets 
would be anti-correlated
with the survey volume.
As carefully discussed in \cite{SE}, a
survey having $\bar{n}_gP_g\simgt 3$ over the range of wavenumbers
considered is close to an optimal design.  

To make the comparison between different survey designs easier, we 
shall fix the total sky coverage of the surveys to
$$
\Omega_{\rm survey} = 300~{\rm deg^2},
$$
for all cases.
We choose to work with three surveys observing at three different
redshift ranges: 
\begin{itemize}
 \item G1: $0.5<z<2$
 \item G2: $2<z<4$
 \item SG: $3.5<z<6.5$
\end{itemize}
where G1 and G2 stands for the ``Ground-based galaxy survey'' 1 and 2,
respectively, while SG stands for the ``Space-based Galaxy survey''. 
Table~1 lists detailed survey parameters for each survey design.

\subsection{G1: Ground-based Galaxy Survey at $0.5<z<2$}

The first survey design, G1, is limited to $0.5<z<2$ for the following
reasons.
One of the reasonable target galaxies from the ground 
would be giant ellipticals or star-forming galaxies because of
their large luminosity.
If spectroscopic observations in
optical wavebands are available, galaxies having either 3727 \AA [OII]
emission lines or $4000$ \AA~ continuum break may be selected, in which
case $z=1.3$ would be the highest redshift bin, as these spectral
line features will move out of the optical wavebands otherwise. 
If spectroscopy in the near infrared band is available, such as 
that proposed by the FMOS instrument on Subaru
telescope, one may select 6563 \AA~ H$\alpha$ emission lines which usually
have the highest equivalent width among the lines in a star-forming
galaxy, in which case a higher redshift, $z\simlt 2$, may be reached. 
Based on these considerations, we consider a survey of $0.5<z<2$ and subdivide
the survey into 3 redshift bins centering at $z=0.75$, $1.25$ and 1.75
with widths $\Delta z=0.5$. While it is currently difficult to
estimate the number density and bias parameters for these galaxies
with any certainty
because we have a limited knowledge of how such galaxies formed within 
the CDM hierarchical clustering scenario, we follow the argument given in
\cite{SE,Blake,Taka03} and assume $\bar{n}_g=0.5 \times 10^{-3}$
$h^3$~Mpc$^{-3}$.
We determine $b_1$ so that it satisfies the condition $\sigma_{8,g}=1$
at a given redshift, where
\begin{equation}
\sigma_{8,g}=b_1\sigma_{8,{\rm
 mass}}\sqrt{1+\frac{2\beta_m}{3}+\frac{\beta_m^2}{5}},
\end{equation}
with $\beta_m=-d\ln D/d\ln (1+z)$ (i.e. 
we do not include the massive neutrino contribution 
when we estimate the fiducial $b_1$). 
Note that 
there is no {\it a priori} reason to believe that 
the r.m.s. fluctuation of the number density of galaxies within
an 8~$h^{-1}$~Mpc sphere should be unity; this condition is rather
motivated by observations, and it does not have to be true for
arbitrary population of galaxies.
Nevertheless, this approach seems to provide reasonable values
for $b_1$, and also it makes it easier to compare our results
with the previous work that used the same recipe \cite{SE,taka05}.
The values of $k_{\rm max}$ in Table \ref{tab:survey} are
computed by $\sigma_{\rm mass}(R,z)=0.5$, 
where $\sigma_{\rm mass}(R,z)$ is the r.m.s.
mass fluctuation in a sphere of radius corresponding to $k_{\rm max}$,
$R=\pi/(2k_{\rm max})$ (see Sec.~\ref{distort}).

\subsection{G2: Ground-based Galaxy Survey at $2<z<4$}

The second design, G2, probes higher redshifts than G1 by 
observing different tracers. The primary candidates from the ground 
in this redshift range would be
Lyman-break galaxies or Lyman-$\alpha$ emitters, which 
are accessible from a deep
survey of 8-m class telescopes in optical wavebands. This
type of survey has been proposed by the 
Hobby--Ebery Telescope Dark Energy eXperiment (HETDEX) \cite{hetdex} and 
Wide Field Multiple Object Spectrograph (WFMOS)
collaborations \cite{wfmos}.  
To make the comparison easier, we shall assume the same number density
as for G1, $0.5\times 10^{-3}$~$h^3$~Mpc$^{-3}$, for G2.
This number corresponds to 4500 galaxies per square degrees, or
1.25 galaxies per square arcminutes for the surface density.
We subdivide the redshift range of G2 into 4 bins, 2.25, 2.75, 3.25 and
3.75, and again determine the bias
parameters by imposing $\sigma_{8,g}(z)=1$ at center redshifts of each
bin.

\subsection{SG: Space-based Galaxy Survey at $4<z<6$}

The third design, SG, is a space-based observation which targets galaxies at
even higher redshifts, $4\simlt z \simlt 6$.
The useful line features will be redshifted into infrared, which makes
such high-$z$ galaxies accessible only from space.
We determine the survey parameters on the basis of 
the Cosmic Inflation Probe (CIP) 
mission \cite{cip}, 
one of the nine studies selected by NASA to investigate
new ideas for future mission concepts within its Astronomical Search 
for Origins Program.
The CIP is a slitless-grating survey in the near infrared, $2.5$--$5$ $\mu$m, 
which detects H$\alpha$ emission lines in star-forming galaxies
at these redshifts.
Being up in space with low background, CIP can achieve a superb 
sensitivity in infrared.  We assume the number density of $5\times
10^{-3}$~$h^3$~Mpc$^{-3}$ \cite{cip}, which is larger by one order 
magnitude than that by the ground based surveys. 
This number density may be partly justified by
the fact that the Lyman break galaxies or the Lyman-$\alpha$ emitters
show the similar number density at these redshifts as implied from a
deep imaging survey \cite{Ouchi}, and most of such galaxies are very likely to
exhibit an even stronger H$\alpha$ emission line. 
For this survey,
we
assume the bias parameters, $b_1=4.5$, $5$ and $5.5$ for redshift slices
of $z=4$, $5$ and $6$ with redshift width $\Delta z=1$, respectively. 
The bias parameters for this survey have been determined using 
a different method. We used the mass-weighted mean halo bias
above a certain minimum mass, $M_{\rm min}$. The minimum mass was found
such that the number density of dark matter halos above 
$M_{\rm min}$ should match the assumed number density of galaxies, 
$\int_{M_{\rm min}}^\infty \! \!dM~dn/dM = 5\times
10^{-3}$~$h^3$~Mpc$^{-3}$. Therefore, we basically assumed that
each dark matter halo above $M_{\rm min}$ hosts one H$\alpha$ emitter 
on average. One may improve this model by using the Halo Occupation
Distribution model, at the expense of increasing the number of 
free parameters.

Note that we chose these survey designs not to say these 
are the optimal designs for doing cosmology with high-$z$ surveys,
but rather to show how well these planned surveys can constrain the neutrino
and inflationary parameters.
We are hoping that our results provide some useful information
in designing high-$z$ galaxy surveys.

\begin{table*}
\begin{center}
\begin{tabular}{l|c \colskip ccc \colskip 
c
@{\hspace{1.em}}
c
@{\hspace{1.em}}
c
@{\hspace{1.em}}
c
@{\hspace{1.em}}
c
@{\hspace{1.em}}
c}
\hline\hline
&&  $N^{\rm nr}_{\nu,{\rm fid}}=1$ & $N^{\rm nr}_{\nu,{\rm fid}}=2$ 
& $N^{\rm nr}_{\nu,{\rm
 fid}}=3$ & \\
Survey & $f_\nu(m_{\nu,{\rm tot}}~ {\rm eV})$ & $N^{\rm nr}_\nu$ 
& $N^{\rm nr}_\nu$ 
& $N^{\rm nr}_\nu$ 
& $n_s$ & $\alpha_s$ &
$\Omega_{\rm m}$ & $\ln\delta_{\cal R}$
& $\ln\Omega_{\rm m}h^2$ & $\ln\Omega_{\rm b}h^2$
\\
\hline 
Planck alone  & 
-- & -- & -- & -- & 0.0062 & 0.0067 & 0.035 & 0.013 & 0.028 & 0.011
\\
G1
&0.0045(0.059) & 0.31 &
0.64 &
1.1 & 
0.0038 &
0.0059 &
0.0072 &
0.0099 &
0.0089 &
0.0075 
\\
G2 
&0.0033(0.043) & 0.20 &
0.49 &
0.90 & 
0.0037 &
0.0057 &
0.0069 &
0.0099 &
0.0086 &
0.0072 
\\
SG 
&0.0019(0.025) & 0.14 &
0.40 &
0.80 & 
0.0030 &
0.0024 &
0.0041 &
0.0090 &
0.0055 &
0.0050 
\\
All (G1+G2+SG)
&0.0018(0.024) & 0.091 &
0.31 &
0.60 & 
0.0026 &
0.0023 &
0.0030 &
0.0089 &
0.0043 &
0.0048 
\\
\hline
\end{tabular}
\end{center}
\caption{\label{tab:fnu005} 
The projected 68\% error on the cosmological parameters
from Planck's CMB data alone 
(the 1st row) and the high-$z$ galaxy survey data combined with the
Planck data (from the 2nd to 5th rows). 
The quoted error for a given parameter includes 
marginalization over the other parameter uncertainties.
 Note that the
 values with and without parenthesis in the 2nd column are the errors
 for $f_\nu$ and $m_{\nu, {\rm tot}}$ (eV), respectively. 
The fiducial values for the neutrino parameters are
 $f_{\nu,{\rm fid}}=0.05$ (the neutrino mass fraction; Eq.~[\ref{eq:fnu}]) 
and $N^{\rm nr}_{\nu,{\rm fid}}=3$ (the number of non-relativistic
 neutrinos), while
the fiducial values for the other parameters are given 
in Sec.~\ref{model}. 
We also vary the fiducial value of $N^{\rm nr}_{\nu,{\rm fid}}$ from 1 to 2 and 3 
when quoting
the projected errors for $N^{\rm nr}_{\nu}$ with $f_{\nu,{\rm
 fid}}=0.05$ being fixed,  as indicated
in the 3rd and 5th columns.
}
\end{table*}
\begin{table*}
\begin{center}
\begin{tabular}{l|c \colskip ccc \colskip 
c
@{\hspace{1.em}}
c
@{\hspace{1.em}}
c
@{\hspace{1.em}}
c
@{\hspace{1.em}}
c
@{\hspace{1.em}}
c}
\hline\hline
&&  $N_{\nu,{\rm fid}}=1$ & $N_{\nu,{\rm fid}}=2$ & $N_{\nu,{\rm
 fid}}=3$ & \\
Survey & $f_\nu(m_{\nu,{\rm tot}}~ {\rm eV})$ & $N^{\rm nr}_\nu$ 
& $N^{\rm nr}_\nu$ 
& $N^{\rm nr}_\nu$ 
& $n_s$ & $\alpha_s$ &
$\Omega_{\rm m}$ & $\ln\delta_{\cal R}$
& $\ln\Omega_{\rm m}h^2$ & $\ln\Omega_{\rm b}h^2$
\\
\hline 
G1
&0.0044(0.058) & 2.2 &
7.1 &
14 & 
0.0037 &
0.0059 &
0.0069 &
0.0099 &
0.0085 &
0.0073 
\\
G2 
&0.0033(0.043) & 2.1 &
7.1 &
13 & 
0.0036 &
0.0058 &
0.0059 &
0.0098 &
0.0075 &
0.0066 
\\
SG 
&0.0021(0.028) & 1.9 &
6.4 &
13 & 
0.0028 &
0.0021 &
0.0034 &
0.0090 &
0.0048 &
0.0048 
\\
All (G1+G2+SG)
&0.0019(0.025) & 1.1 &
3.7 &
7.4 & 
0.0021 &
0.0016 &
0.0017 &
0.0087 &
0.0030 &
0.0045 
\\
\hline
\end{tabular}
\end{center}
\caption{\label{tab:fnu001} Same as in the previous table, but for 
the smaller fiducial neutrino mass fraction, $f_{\nu,{\rm fid}}=0.01$.}
\end{table*}
%

\section{Parameter Forecast: Basic Results}
\label{results}

Tables~\ref{tab:fnu005} and \ref{tab:fnu001} summarize the basic
results of our forecasts for the cosmological
parameters from the high-$z$ galaxy redshift surveys
combined with the Planck data.
Each column shows the projected 1-$\sigma$ error on a particular
parameter, marginalized over the other parameter uncertainties.
The 1st row in Table~\ref{tab:fnu005} 
shows the constraints from the Planck data alone, while
the other rows show the constraints from the Planck data combined with
each of the high-$z$ galaxy surveys outlined in the previous section.
The final row shows the constraints from all the data combined.
The difference between these two Tables is the fiducial value for
$f_\nu=\Omega_\nu/\Omega_{\rm m}$: Table~\ref{tab:fnu005} uses
$f_\nu=0.05$, whereas Table~\ref{tab:fnu001} uses
a lower value, $f_\nu=0.01$, as the fiducial value.
In addition, in each Table the fiducial value for the number 
of non-relativistic neutrino species, $N_\nu^{\rm nr}$, is also
varied from $N_\nu^{\rm nr}=1$ to 2 to 3.
Therefore, in Table~\ref{tab:fnu005} the fiducial mass of individual
non-relativistic species changes from $m_\nu=0.66$ to 0.33 to 0.22~eV,
whereas in Table~\ref{tab:fnu001} it changes from
$m_\nu=0.13$ to 0.066 to 0.044~eV for $N_\nu^{\rm nr}=1$, 2, and 3,
respectively.
It is also worth showing how the parameter errors are correlated with
each other, and we give the parameter correlations in
Table~\ref{tab:coeff} for the case of SG combined with Planck
in Appendix~\ref{correl}.

\subsection{Neutrino Parameters}

\begin{figure}
  \begin{center}
    \leavevmode\epsfxsize=8.5cm \epsfbox{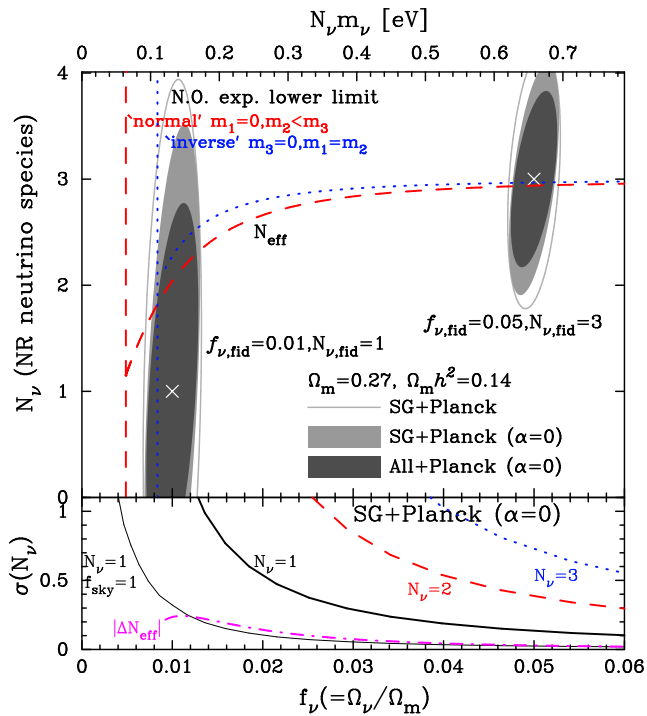}
  \end{center}
\vspace{-2em}
\caption{ {\em Upper panel}: Projected 68$\%$ error ellipses in the
neutrino parameter, ($f_\nu$-$N^{\rm nr}_\nu$) plane, 
 expected from the high-$z$ galaxy survey data combined with the Planck data 
(see Table \ref{tab:survey} for the survey definition).  
The two fiducial models for $f_\nu$ and $N^{\rm nr}_\nu$ are 
considered: the left contours assume 
$(f_{\nu,{\rm fid}}, N^{\rm nr}_{\nu,{\rm fid}})=(0.01,1)$, while 
the right contours assume 
$(f_{\nu,{\rm fid}}, N^{\rm nr}_{\nu,{\rm fid}})=(0.05,3)$.
The outer thin lines and the middle light-gray contours are 
the forecasts 
for SG (the space-based mission at $3.5<z<6.5$) 
plus Planck, 
without and with a prior on the running spectral index,
$\alpha_s=0$, respectively.
The innermost, dark gray contours show the forecasts when
all the galaxy surveys (two ground-based surveys and SG) and Planck are 
combined. 
The vertical dashed and dotted lines
show the lower limits on $f_\nu$ implied from the neutrino oscillation
experiments assuming the normal and inverted mass hierarchy models,
respectively.  
The dashed and dotted curves then show the effective number of
non-relativistic neutrino species, $N_{\rm eff}$,
for  the two hierarchy models (see Eq.~[\ref{eq:neff}] for the definition).
{\em Lower panel}: The projected 68\% 
on $N^{\rm nr}_{\nu}$ as a function of the fiducial value of $f_\nu$.
The thick solid, dashed, and dotted lines use 
the fiducial values 
of $N^{\rm nr}_{\nu,{\rm fid}}=1$, 2, and 3, respectively. 
The dot-dashed curve shows the difference between $N_{\rm eff}$ 
for the normal and inverted mass hierarchy models.  
The leftmost thin solid line shows the error expected from
a hypothetical full-sky SG survey for $N^{\rm nr}_{\nu,{\rm fid}}=1$.
}\label{fig:fnu-Nnu}
\end{figure}

In this paper, we are particularly interested in the capability
of future high-$z$ redshift surveys to constraining 
the neutrino parameters, $f_\nu$ and $N^{\rm nr}_\nu$, as well as
the shape of the primordial power spectrum, the tilt ($n_s$) and the running
spectral index ($\alpha_s$).  First, we study the neutrino parameters.

The upper panel of Figure \ref{fig:fnu-Nnu} shows error ellipses in the
subspace of ($f_\nu$, $N^{\rm nr}_\nu$).
Two ``islands'' show two different fiducial models: the left island is 
$(f_\nu, N^{\rm nr}_\nu)=(0.01, 1)$, while
the right island is $(0.05, 3)$.
We find that the errors on $f_\nu$ and $N^{\rm nr}_\nu$ are only weakly
degenerate with each other, implying that the constraints on the two
parameters come from different regions of $P(k)$ in $k$-space,
which can be seen more clearly from Figure~\ref{fig:pk}.

\subsubsection{Total Neutrino Mass}

\begin{figure}
  \begin{center}
    \leavevmode\epsfxsize=9.cm \epsfbox{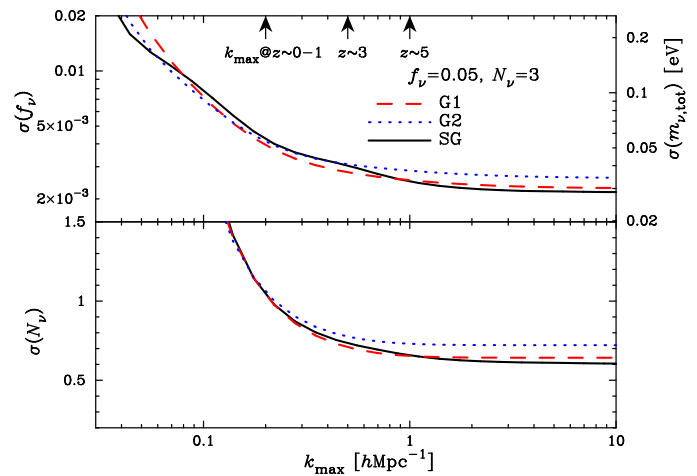}
  \end{center}
\vspace{-2em}
\caption{
The projected 68\% error on $f_\nu$ (upper panel) 
and $N^{\rm nr}_\nu$ (lower panel) against the maximum wavenumber,
$k_{\rm max}$,  assuming the information in the linear 
power spectrum at $k\le k_{\rm max}$ can be
used in the Fisher matrix analysis.
The arrows in the above $x$-axis indicate our nominal $k_{\rm max}$ 
used in the analysis at each redshift 
(see Table \ref{tab:survey}).  
Note that $f_{\nu,{\rm fid}}=0.05$ and $N_{\nu,{\rm fid}}^{\rm nr}=3$
are assumed.
}\label{fig:kmax}
\end{figure}

As we have shown in Sec.~II, galaxy surveys constrain 
the total mass of non-relativistic neutrinos by measuring 
the overall suppression of power at small scales compared with 
the scales larger than the neutrino free-streaming length,
$\Delta P(k)/P(k)\simeq -8f_\nu$
[Eq.~(\ref{eq:overallsuppression})].
Tables \ref{tab:fnu005} and
\ref{tab:fnu001} and Figure \ref{fig:fnu-Nnu} (the widths of the 
error ellipses show the accuracy of constraining
the total neutrino mass,
$N_\nu^{\rm nr}m_\nu$) show that
the high-$z$ galaxy surveys can provide very tight constraints on the
total neutrino mass, $m_{\nu, {\rm tot}}=N_\nu^{\rm nr}m_\nu$, 
and the constraint improves steadily by going to higher redshifts 
for a given survey area.
The projected error (assuming $f_\nu=0.05$) improves from 
$\sigma(m_{\nu, {\rm tot}})= 0.059$ to 0.025~eV for G1 to SG.
(The constraint on $m_{\nu,{\rm tot}}$ is very similar for $f_\nu=0.01$.)
This is because the suppression rate of the amplitude of the linear power spectrum
on small scales can be precisely measured by the galaxy survey
when combined with the tight constraint on the amplitude of the spectrum
on large scales from Planck (also see Table~\ref{tab:noCMB} and
Table~\ref{tab:coeff} in Appendix~\ref{correl}).  
The steady improvement at higher redshifts is simply because 
surveys at higher redshifts can be used to probe smaller spatial scales
(i.e., larger $k_{\rm max}$; see Table~I).
We find that it is crucial to increase $k_{\rm max}$ as much as possible
in order to improve the constraints on the neutrino parameters.
Figure \ref{fig:kmax} shows how one can reduce the
errors on $f_\nu$ and $N^{\rm nr}_\nu$ by increasing $k_{\rm
max}$.

These results are extremely encouraging.
If we {\em a priori} assume three-flavor neutrinos in
compatible with the neutrino oscillation experiments, then 
it is likely that one can determine the sum of neutrino masses 
using the high-$z$ galaxy redshift surveys, as
the current lower bounds implied from the neutrino oscillation experiments
are $m_{\nu, {\rm tot}}\simgt 0.06$ and $0.1\,$~eV for the normal
and inverse mass hierarchies, respectively.  
It should also be noted that a
detection of the total mass in the range of $m_{\nu,{\rm tot}}<0.1\,$~eV
gives an indirect evidence for the normal mass hierarchy, thereby
resolving the mass hierarchy problem.

Our results may be compared with the previous work \cite{HET}, where
$\sigma(m_{\nu, {\rm tot}})\approx 0.3\,$eV ($1\sigma$) was obtained;
therefore, our errors are smaller than theirs 
by a factor of 5--10, even though the
survey volume that we assumed is larger only by a factor of 1.5--3 than
what they assumed. 
What drives the improvement? 
There are two reasons. The first reason is because we consider 
high-$z$ galaxy surveys, while  \cite{HET} considered low-$z$ surveys,
such as the Sloan Digital Sky Survey, which
suffer from much stronger non-linearity.
We are therefore using the information on the power spectrum
down to larger wavenumbers (i.e., $k_{\rm max}$ is larger).
Figure~\ref{fig:kmax} shows that all the galaxy surveys have essentially
equal power of constraining the neutrino parameters, 
when the information up to the same $k_{\rm max}$ is used.
However, one cannot do this for low-$z$ surveys because of 
strong non-linearity. 
As long as we restrict ourselves to the linear regime, a
higher redshift survey is more powerful in terms of constraining the
neutrino parameters.
Interestingly, the error on the neutrino parameters appears to be
saturated at $k_{\rm max}\sim 1~{\rm Mpc}^{-1}$; thus, 
the space-based mission, SG, is already nearly optimal for constraining the
neutrino parameters
for the survey parameters (especially the number density and bias parameters of
sampled galaxies that determine the shot noise contribution to limit the
small-scale measurements).
The second reason 
is because our parameter forecast uses the full 2D
information in the redshift-space power spectrum (see
Eq. (\ref{eqn:ps})) that includes effects of the cosmological distortion
and the redshift-space  distortion due to peculiar velocity.
These effects are very useful in breaking parameter degeneracies.  
Table \ref{tab:noCMB} shows that the
constraint on $m_{\nu, {\rm tot}}$ would be significantly degraded 
if we did not include the
distortion effects. In particular, ignoring the information on the
redshift-space distortion,
which is consistent with the analysis of 
\cite{HET}, leads to a similar-level constraint on
$m_{\nu,{\rm tot}}$
as theirs.
The inclusion of the redshift-space distortion helps
break degeneracy between the power spectrum amplitude
and the galaxy bias, which in turn helps determine the small-scale 
suppression due to the
neutrino free-streaming.

The projected error on
the neutrino total mass might also 
depend on the fiducial value of $\Omega_{\rm m}h^2$, 
as the effect of neutrinos 
on the power spectrum depends on 
$f_\nu\propto m_{\nu, {\rm tot}}/(\Omega_{\rm m}h^2)$
[Eq.~(\ref{eq:fnu})] \cite{HET}.
For a given $f_\nu$, a variation in $\Omega_{\rm m}h^2$ changes
$m_{\nu, {\rm tot}}$.
Now that $\Omega_{\rm m}h^2$ has been constrained accurately
by WMAP, however, we find that our results are not very sensitive
to the precise value of $\Omega_{\rm m}h^2$.
We have repeated our analysis for  $\Omega_{\rm m}h^2=0.1$,
the $2\sigma$-level lower bound from the WMAP results \cite{WMAP},
and found very similar results.

\subsubsection{Number of Non-relativistic Neutrino Species}

\begin{table*}
\begin{center}
\begin{tabular}{l|c@{\hspace{1em}}c  @{\hspace{8em}} 
c@{\hspace{1em}}c@{\hspace{1em}}c}
\hline\hline
& \multicolumn{2}{c}{\hspace{-0em}+full Planck ($@ k_0=0.05~$Mpc$^{-1}$)}
& \multicolumn{3}{c}{@ $k=k_{\rm
 pivot}$} 
\\
& \hspace{5em}$n_s$ & $\alpha_s$ 
& $k_{0,{\rm pivot}}$ (Mpc$^{-1}$)& $n_s$ & $\alpha_s$ \\
\hline 
Planck alone & \hspace{5em}0.0062 & 0.0067 & -- & -- & --\\
G1 & \hspace{5em}0.0038 & 0.0059 &  0.030 & 0.086 & 0.035\\
G2 & \hspace{5em}0.0037 & 0.0057 &  0.18 & 0.018 & 0.025\\
SG &\hspace{5em}0.0030 & 0.0024 & 0.48 & 0.0033 & 0.0070\\
\hline
\end{tabular}
\end{center}
\caption{\label{tab:k0pivot} 
Projected 68\% errors on the parameters that characterize
 the shape of the primordial power spectrum, 
 the tilt ($n_s$) and the running index $(\alpha_s)$.
 The left block is the same as the 6th and 7th columns in
 Table~\ref{tab:fnu005}, which combines the galaxy survey data with
 the CMB data from Planck, while the right block shows the constraints
 when the CMB information on $n_s$ and $\alpha_s$ are not used. 
 The left block lists the constraints on $n_s$ and $\alpha_s$ using 
 $k_0=0.05~{\rm Mpc}^{-1}$ [see Eq.~(\ref{eq:pknorm})]
which was chosen such that
 the Planck data would yield the best constraints.
 The 1st column in the right block shows the pivot wavenumber
 at which the errors on $n_s$ and $\alpha_s$ are uncorrelated for a given
 galaxy survey, 
 and the 2nd and 3rd columns show the constraints on $n_s$ and
 $\alpha_s$ at the
 pivot wavenumber, respectively. 
 The space-based galaxy survey at $3.5<z<6.5$, SG, on its own
 yields better constraints on $n_s$ and comparable constraints on
 $\alpha_s$ compared to Planck alone, when evaluated  at its pivot wavenumber.
 Note that
 $f_{\nu,{\rm fid}}=0.05$ and $N^{\rm nr}_{\nu,{\rm fid}}=3$ are assumed. }
\end{table*}

Galaxy surveys could also be used to determine
the {\it individual} mass of non-relativistic neutrinos, $m_\nu$.
As we have shown in Sec.~II, galaxy surveys can constrain 
$m_\nu$ by determining the free-streaming scale, 
$k_{\rm fs}(z)\propto m_\nu$ [Eq.~(\ref{eq:kfs*})],
 from distortion of the shape of the galaxy power spectrum
near $k_{\rm fs}(z)$.

Neutrino oscillation experiments have provided tight limits
on the mass square differences between neutrino mass eigenstates as
$|m_2^2-m^2_1|\simeq 7\times 10^{-5}~$eV$^2 $
and $|m^2_3-m^2_2|\simeq 3\times 10^{-3}~$eV$^2$ 
where $m_i$ denotes mass of the $i$-th mass eigenstates. 
We model a family of possible models by the largest neutrino mass,
$m_\nu$, motivated by the fact that 
 structure formation is sensitive to most massive species. 
We may define the effective number of non-relativistic neutrino
species as
\begin{equation}
 N_{\rm eff}^{\rm nr}\equiv 1+\frac{m_i}{m_\nu}+\frac{m_j}{m_\nu},
\label{eq:neff}
\end{equation}
which continuously varies between $1\leq N_{\rm eff}^{\rm nr}\leq 3$.
The total neutrino mass is given by $m_{\nu,{\rm tot}}=
 N_{\rm eff}^{\rm nr}m_\nu$.
We then consider two neutrino mass hierarchy models (we shall assume, by
convention, that $m_2\ge m_1$):

\begin{itemize}
\item {\it Normal mass hierarchy}: $m_\nu=m_3$

In this model, $m_3$ is assumed to be the largest mass, and
$m_1<m_2<m_3$ 
are allowed.
When $m_2, m_1\ll m_3$ (the extreme case $m_1=0~$eV), $N_{\rm eff}^{\rm
       nr}\simeq 1$. 
On the other hand, when $m_1$ and $m_2$ are comparable to 
the mass difference between $m_2$ and $m_3$, we have
$N_{\rm eff}^{\rm nr}\simeq 3$ (i.e., three masses are nearly
degenerate).
Hence, the ``normal mass hierarchy'' model allows $N_{\rm eff}^{\rm nr}$
to vary in the full parameter space,  $1\leq N_{\rm eff}^{\rm nr}\leq 3$.

\item {\it Inverted mass hierarchy}, $m_1\sim m_2=m_\nu$
 
In this model, $m_3$ is assumed to be the smallest mass, and
$m_3<m_1<m_2$. 
A peculiar feature of this model is that  $N_\nu^{\rm nr}$
cannot be less than 2: 
when $m_3=0$, the possible solutions allowed by the neutrino
oscillation experiments are $m_2\simeq 0.055$~eV and 
$m_1\simeq 0.047$~eV or $m_{\nu,{\rm tot}}\simeq 0.1$~eV; thus, 
$m_1$ and $m_2$ must be very similar, giving $N_\nu^{\rm nr}\simeq 2$.
Again, when $m_3$ is comparable to 
the mass difference between $m_2$ and $m_3$, all three masses are degenerate,
$N_{\rm eff}^{\rm nr}\simeq 3$.
Hence, the ``inverted mass hierarchy'' model allows $N_{\rm eff}^{\rm nr}$
to vary only in the limited parameter space,  $2\leq N_{\rm eff}^{\rm nr}\leq 3$.
\end{itemize}

In the upper panel of Figure \ref{fig:fnu-Nnu}, we show  $N_{\rm eff}^{\rm nr}$
for the normal mass hierarchy model (dashed line) and for 
the inverted mass hierarchy model (dotted line). One can see that
the two models are indistinguishable (all masses are degenerate)
for  $f_\nu\gtrsim 0.02$.

How do we constrain  $N_\nu^{\rm nr}$?
We measure $m_{\nu,{\rm tot}}$ from the overall suppression of
power at small scales, as described in the previous section.
Then we measure $m_\nu$ from the ``break'' of the power spectrum
caused by the free-streaming scale, $k_{\rm fs}(z)$.
The number of non-relativistic neutrinos is finally constrained
as $N_\nu^{\rm nr}=m_{\nu,{\rm tot}}/m_\nu$, which tells us about 
the neutrino mass hierarchy.
In the 3rd to 5th columns in Table~\ref{tab:fnu005}
we show the projected error on $N_\nu^{\rm nr}$ from
high-$z$ galaxy surveys combined with the Planck data,
assuming the fiducial neutrino mass fraction of $f_\nu=0.05$. 
As the error depends very much on
the fiducial value of $N_\nu^{\rm nr}$, we explore three different
fiducial values, $N_{\nu, {\rm fid}}^{\rm nr}=1$, 2 and 3.
As we have described above, however, the first two fiducial values,
$N_{\nu,{\rm fid}}^{\rm nr}=1$ and 2, are inconsistent with the neutrino
oscillation experiments if $f_\nu=0.05$ (all the masses must be nearly degenerate); 
thus, only the 5th column is actually realistic if there is no
sterile neutrino.
We find that $N_\nu^{\rm nr}$ is going to be difficult to
constrain: even when we combine all the high-$z$ galaxy surveys,
G1, G2 and SG, the projected error is $\Delta N_\nu^{\rm nr}=0.6$.
This implies that it is not possible to discriminate between
$N_\nu^{\rm nr}=2$ and 3 at more than 2-$\sigma$, while one can
reject $N_\nu^{\rm nr}=1$.
When the fiducial value of $f_\nu$ is small enough to allow for
$N_\nu^{\rm nr}\leq 2$, $f_\nu=0.01$, the constraints 
are too weak to be useful (see the 3rd to 5th columns
in Table~\ref{tab:fnu001}).

As explained above, 
if the total neutrino mass is larger than 0.2\,eV, 
$N^{\rm nr}_{\rm eff}\simeq 3$ is expected
from the neutrino oscillation experiments. 
The lower panel of Figure \ref{fig:fnu-Nnu} shows the projected error on
$N_\nu^{\rm nr}$ for the SG survey. A model with
$N^{\rm nr}_\nu=3$ can be detected  at more than $1\sigma$ level only if
$f_\nu\simgt 0.04$ $(m_{\nu, {\rm tot}}\simgt 0.52\,$eV).  Nevertheless, 
it should be
noted that exploring the constraint on $N^{\rm nr}_\nu$ with future
surveys is extremely
important  because any finding of 
a model with $N^{\rm nr}_\nu\ne 3$ in this range of $f_\nu$ 
may provide valuable information for the 
existence of sterile neutrino or new physics. 
The general trend is that the error on $N_\nu^{\rm nr}$ increases
as $m_\nu$ decreases (i.e., the fiducial value of
$f_\nu$ decreases or $N_\nu^{\rm nr}$ increases). 
This is because the neutrino free-streaming scale, $k_{\rm fs}$,
is proportional to
$m_\nu$: when $m_\nu$ is too small, $k_{\rm fs}$ will go out of 
the $k$ range accessible  by galaxy surveys.
A possible way to overcome this obstacle is to enlarge the survey volume
which, in turn, can lower the minimum wavenumber
sampled by the survey. 
The leftmost curve 
shows the projected error on $N_\nu^{\rm nr}$ for
 $N_{\nu,{\rm fid}}^{\rm nr}=1$ from a hypothetical full-sky SG survey
at $3.5<z<6.5$. We find that such a survey will be able to
distinguish between two mass hierarchy models in principle.

\subsection{Shape of the Primordial Power Spectrum}

\begin{figure}
  \begin{center}
    \leavevmode\epsfxsize=8.5cm \epsfbox{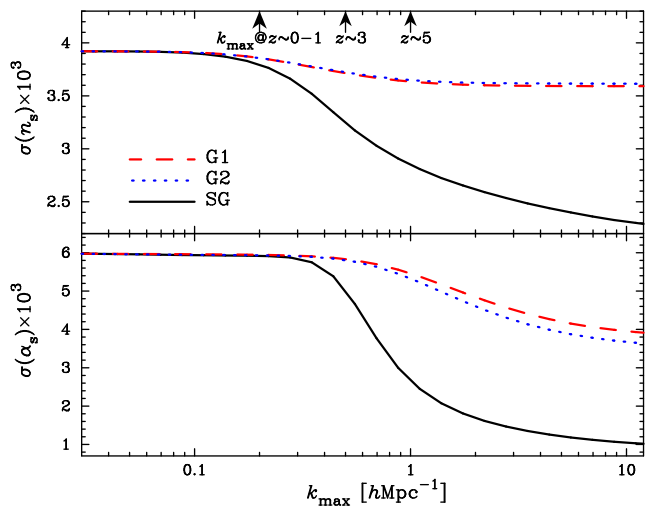}
  \end{center}
\vspace{-2em}
\caption{The projected 68\% error on $n_s$ (upper panel) and $\alpha_s$
 against $k_{\rm max}$, as in Figure \ref{fig:kmax}.
} \label{fig:alpha_kmax}
\end{figure}

\begin{figure*}
  \begin{center}
    \leavevmode\epsfxsize=16.cm \epsfbox{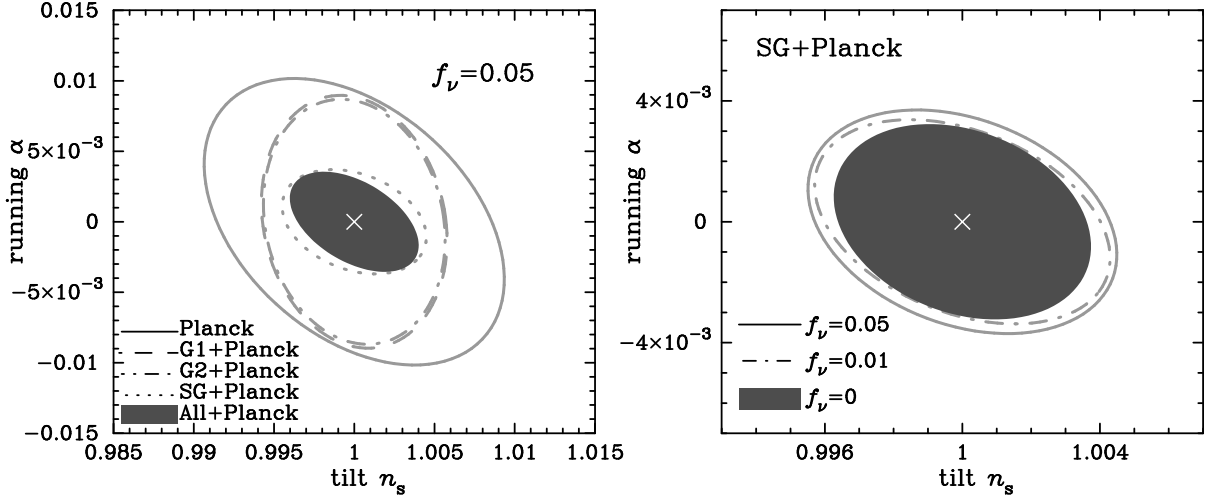}
  \end{center}
\vspace{-2em}
\caption{{\em Left panel}: Projected 68\% error ellipses in the ($n_s$,
 $\alpha_s$) plane from Planck alone (the outermost contour),
 and the high-$z$ galaxy surveys combined with Planck. 
 The dashed and dot-dashed contours are G1 ($0.5<z<2$) and G2 ($2<z<4$), 
 respectively, while the dotted contour is SG ($3.5<z<6.5$).
 The highest redshift survey, SG, provides very
 tight constraints on the shape of the primordial power spectrum.  
 The innermost shaded area shows the constraint from all the galaxy
 surveys and Planck combined.
{\em
 Right panel:} Degradation in the constraints on $n_s$ and $\alpha_s$
 as a function of the non-relativistic neutrino contribution, $f_\nu$, for 
 SG plus Planck. The effect of non-relativistic neutrinos hardly
 affects the constraints on $n_s$ and $\alpha_s$.} \label{fig:alpha}
\end{figure*}

\begin{figure*}
  \begin{center}
    \leavevmode\epsfxsize=16.cm \epsfbox{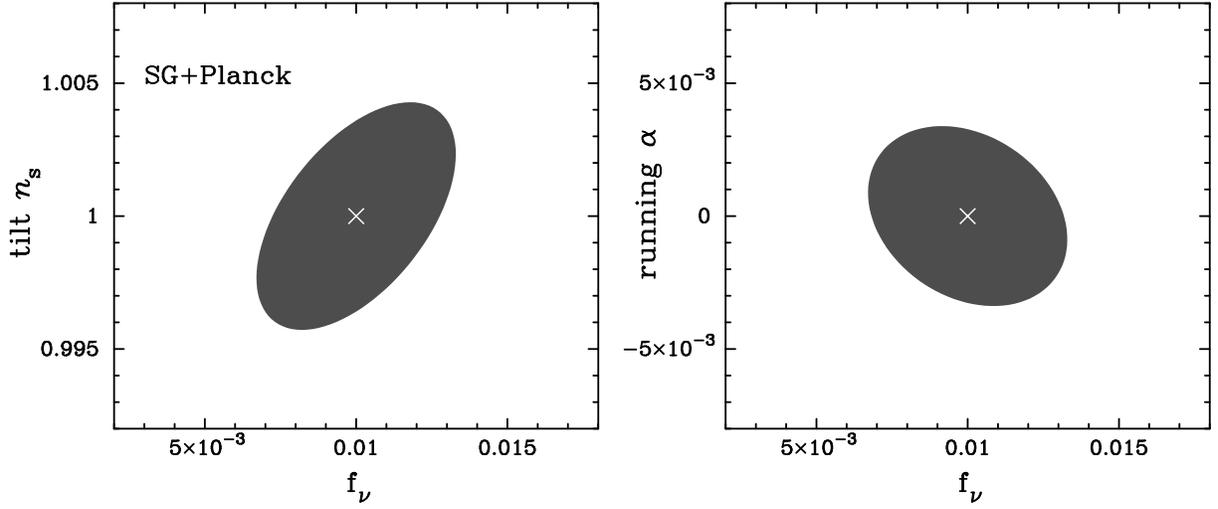}
  \end{center}
\vspace{-2em}
\caption{Projected 68\% error ellipses in the ($f_\nu,n_s$) (left panel)
 and ($f_\nu,\alpha_s$) (right panel) planes, respectively, for  SG
 combined with Planck. Note that the fiducial values of $f_{\nu,{\rm fid}}=0.01$ and $N^{\rm
 nr}_{\nu,{\rm fid}}=1$ are assumed.  
} \label{fig:fnu-ns}
\end{figure*}

\begin{table*}
\begin{center}
\begin{tabular}{l|ccccc \colskip ccccc \colskip ccccc}
\hline\hline
& \multicolumn{5}{c}{\hspace{-2em}No CMB} &
 \multicolumn{5}{c}{\hspace{-2em}No Geometric Distortion}
& \multicolumn{5}{c}{\hspace{-1em}No Redshift Space Distortion}\\
Survey & 
$f_\nu$ & $N^{\rm nr}_\nu$ & $\Omega_{\rm m0}$ & $n_s$ & $\alpha_s$ &
$f_\nu$ & $N^{\rm nr}_\nu$ & $\Omega_{\rm m0}$ & $n_s$ & $\alpha_s$  &
$f_\nu$ & $N^{\rm nr}_\nu$ & $\Omega_{\rm m0}$ & $n_s$ & $\alpha_s$
\\ \hline
G1&0.19&11&0.037&0.31&0.22 & 0.0052 & 1.0 & 0.0074 & 0.0039 & 0.0059 &
 0.017 & 3.8 & 0.0073 & 0.0039 & 0.0060\\
G2&0.082&4.1&0.023&0.10&0.060 & 0.0043 & 0.78 & 0.0073 & 0.0039 &
 0.0057 &
 0.015& 3.0& 0.0070& 0.0038 & 0.0058 \\
SG&0.044&2.6&0.0064&0.057&0.013 & 0.0033& 0.64 & 0.0064 & 0.0038 &
 0.0025 & 0.011 & 2.3 & 0.0048 & 0.0032 & 0.0027\\
\hline
\end{tabular}
\end{center}
\caption{\label{tab:noCMB} 
 Parameter degradation when some information is thrown away from the 
analysis. 
 From left to right, CMB, the geometric distortion, or the redshift
space distortion has been removed from the analysis.
 Note that
 $f_{\nu,{\rm fid}}=0.05$ and $N^{\rm nr}_{\nu,{\rm fid}}=3$ are assumed.}
\end{table*}

\begin{table*}
\begin{center}
\begin{tabular}{l|cccc 
@{\hspace{2em}}cccc@{\hspace{2em}} cccc} 
\hline\hline
& 
\multicolumn{4}{c}{\hspace{-1em}V1N5} & 
\multicolumn{4}{c}{\hspace{-1em}V5N1} & 
\multicolumn{4}{c}{V5N5}  
\\ 
& $f_\nu$ & $N^{\rm nr}_\nu$ & $n_s$ & $\alpha_s$ 
& $f_\nu$ & $N^{\rm nr}_\nu$ & $n_s$ & $\alpha_s$ 
& $f_\nu$ & $N^{\rm nr}_\nu$ & $n_s$ & $\alpha_s$  \\
\hline
G1    & 0.0038 & 0.83(2.0) & 0.0037 & 0.0059 
   & 0.0033 & 0.59(1.4) & 0.0036 & 0.0057 
   & 0.0029 & 0.50(1.3) & 0.0034 & 0.0057 \\
G2    & 0.0024 & 0.71(1.8) & 0.0034 & 0.0052
   & 0.0025 & 0.59(1.4) & 0.0034 & 0.0051
   & 0.0021 & 0.46(1.2) & 0.0028 & 0.0040
\\
SG    & 0.0018 & 0.66(1.6) & 0.0026 & 0.0018
   & 0.0017 & 0.49(1.1) & 0.0023 & 0.0018
   & 0.0016 & 0.42(0.97) & 0.0019 & 0.0013
 \\
\hline
\end{tabular}
\end{center}
\caption{\label{tab:vn} 
 Projected 68\% errors on the neutrino and power spectrum shape
 parameters for advanced survey parameters.
 ``V1N5'' has a factor of 5 larger number density of galaxies than the
 fiducial survey, 
 ``V5N1'' has a factor of 5 larger survey volume,
 and ``V5N5'' has a factor of 5 increase in the both quantities. 
 Note that
 $f_{\nu,{\rm fid}}=0.05$ and $N^{\rm nr}_{\nu,{\rm fid}}=3$ are assumed,
 while  $f_{\nu,{\rm fid}}=0.01$ and $N^{\rm nr}_{\nu,{\rm fid}}=1$ are assumed
for  $N^{\rm nr}_\nu$  in the parentheses.
}
\end{table*}

The amplitude of the primordial power spectrum appears to be 
one of the most difficult parameters to measure very accurately.
Adding the galaxy survey does not help very much: the constraint 
on the amplitude improves only by a factor of 1.5 {\it at
most}, even by combining all the data sets.
(See the 9th column of Table~\ref{tab:fnu005} and \ref{tab:fnu001}.)
This is because the
Planck experiment alone can provide sufficiently tight constraint on the
amplitude: 
Planck allows us to break degeneracy between the
amplitude and the optical depth $\tau$ by measuring the CMB polarization with
high precision (the current accuracy of determining the amplitude,
obtained from the WMAP, is about 10\%).
Adding galaxy surveys does not improve the accuracy of normalization
due to the galaxy bias.
The constraint on the 
amplitude could be further improved by adding 
the weak gravitational lensing data (e.g. see \cite{Hu02,HJ04,TJ}), which
directly measures the dark matter distribution.
Also, the lensing data are actually sensitive to mass clustering in 
the non-linear regime, and thus would be complementary to the 
galaxy surveys probing the linear-regime fluctuations.

The interesting parameter is the running spectral index, $\alpha_s$.
Actually G1 or G2 does not improve the constraint on $\alpha_s$ at all:
the error shrinks  merely by 20\%; however, the space-based survey,
such as CIP, provides a dramatic improvement over the Planck data,
by a factor of nearly 3. This indicates that SG {\it alone} is 
at least as powerful as Planck in terms of constraining $\alpha_s$.
(We shall come back to this point below.)
The driving force for this improvement is the value of the maximum
usable wavenumber for SG, $k_{\rm max}\sim 1~{\rm Mpc}^{-1}$, which is
substantially greater than that for G1, $k_{\rm max}\sim 0.2~{\rm Mpc}^{-1}$,
and that for G2, $k_{\rm max}\sim 0.5~{\rm Mpc}^{-1}$.
Our study therefore indicates that one needs to push $k_{\rm max}$ 
at least up to $k_{\rm max}\sim 1~{\rm Mpc}^{-1}$ in order to achieve
a significant improvement in the constraint on $\alpha_s$, for the
survey parameters (the number density and bias parameters of sampled galaxies) 
we have considered.
This can be also clearly found from Figure \ref{fig:alpha_kmax}. 

On the other hand, the improvement on the tilt, $n_s$, at 
$k_0=0.05~{\rm Mpc^{-1}}$ is similar for G1, G2, and SG: 
from a factor of 1.5 to 2. The interpretation of this result is, however,
complicated by the fact that the actual constraint depends very much
on the value of the pivot scale, $k_0$, at which $n_s$ is defined.
The current value, $k_0=0.05~{\rm Mpc}^{-1}$, was chosen such that
the Planck data would provide the best constraint on $n_s$.
On the other hand, as the galaxy survey data probe fluctuations
on the smaller spatial scales (larger $k$), the optimal pivot wavenumber
for the galaxy surveys should actually be larger than $0.05~{\rm
Mpc}^{-1}$: the optimal pivot wavenumbers for G2 and SG are $k_{0\rm pivot}=0.18$ 
and $0.48~{\rm Mpc^{-1}}$, respectively,
where $k_{0\rm pivot}$ was computed such that
the covariance between $n_s(k_{0\rm pivot})$ and $\alpha_s$
should vanish at $k_{0\rm pivot}$ and the two parameters would be statistically
independent (see \cite{HET} for more discussion on this issue;
see also \cite{SE} for the similar method for constraining the
dark energy equation of state at pivot redshift). 
Table \ref{tab:k0pivot} lists $k_{0\rm pivot}$ for G1, G2 and SG,
and the errors on $n_s(k_{0\rm pivot}) $ and $\alpha_{s,\rm pivot}$.
Note that we do not use the CMB information on $n_s$ and $\alpha_s$ to
derive $k_{0\rm pivot}$ (but include the CMB information on the other
parameters). 
This Table therefore basically shows how the galaxy survey data alone 
are sensitive to the shape of the primordial 
power spectrum. The striking one is SG: the errors on 
$n_s(k_{0\rm pivot})$ and $\alpha_s$ are 0.0033 and 0.0070, respectively,
which should be compared with those from the Planck data alone,
0.0062 and 0.0067. Therefore, SG alone is at least as powerful as
Planck,  
in terms of constraining the shape of the
primordial power spectrum.

The correct interpretation and summary of these results is the following.
The Planck data alone cannot constrain the value of $n_s$ very well
at small spatial scales (the uncertainty diverges at larger $k_0$), 
and the galaxy survey data alone cannot do so at large spatial scales
(the uncertainty diverges at smaller $k_0$). However, when two data sets 
are combined, {\it the accuracy in determining $n_s$ becomes nearly uniform
at all spatial scales}, and the constraint becomes nearly independent of 
a particular choice of $k_0$. This is in fact a huge improvement of the
situation, which may not be seen very clearly from just an improvement of
$n_s$ defined at a particular $k_0$. This is probably 
best represented by the constraint
on $\alpha_s$ we described above: the significant reduction in the uncertainty
in $\alpha_s$ for SG indicates that SG in combination with Planck has nearly
uniform sensitivity to the shape of the primordial power spectrum
from CMB to galaxy scales. This is exactly what one needs for 
improving constraints on inflationary models.

The left panel of Figure \ref{fig:alpha} summarizes the constraints 
on $\alpha_s$ and $n_s$ at $k_0=0.05~{\rm Mpc^{-1}}$. This figure
shows the projected error ellipses in the
($n_s$, $\alpha_s$) subspace. The overall improvement on the parameter
constraint in the 2D sub-space is quite impressive, and it clearly
shows the importance of high-$z$ galaxy surveys for improving
constraints on the shape of the primordial power spectrum.

Yet, one might be worried about the presence of non-relativistic neutrinos
degrading the constraints, $\alpha_s$ in particular, as the neutrinos
might mimic the effect of a negative $\alpha_s$ by suppressing 
the power more at smaller spatial scales.
The right panel of Figure \ref{fig:alpha} basically shows that
there is no need to worry: the constraints on $n_s$ and $\alpha_s$ 
are hardly affected by the non-relativistic neutrinos for $f_\nu=0-0.05$
(the errors on $n_{\rm s}$ and $\alpha_{\rm s}$ are degraded less
than by
$\sim 10\%$).
This is because the inflation parameters and the non-relativistic
neutrinos change the shape of the galaxy power spectrum in modestly
different ways, as explicitly demonstrated in the lower panel of Figure
\ref{fig:pk}, and  
the high-$z$ galaxy surveys are capable of discriminating these effects.
Figure \ref{fig:fnu-ns} shows the error ellipses in
the sub-spaces of $(f_\nu, n_{\rm s})$ (left panel) and ($f_\nu, \alpha_{\rm
s}$) (right panel).  
We find that the correlation between $n_{\rm s}$ and $f_\nu$ is modest
(the correlation coefficient defined by Eq.~(\ref{eqn:coeff})
is $0.55$; see also Table~\ref{tab:coeff}), while the correlation
between $\alpha_{\rm s}$ and $f_\nu$ is weaker. 
The most important result from this study is therefore that
the 2D joint marginalized constraint on inflationary parameters,
$n_{\rm s}$ and $\alpha_{\rm s}$, is hardly affected by the presence of 
non-relativistic neutrinos.

\subsection{Information from Geometric and Redshift Space Distortion}

In Table \ref{tab:noCMB} we summarize what happens when we throw away
some information from our analysis.
Without CMB information, the errors on  the neutrino parameters
inflate significantly by more than an order of magnitude, 
while the error on the matter density
parameter, $\Omega_{\rm m}$, is still comparable to or better than that from
Planck alone. (Why this is so is explained in the next paragraph.)
The errors on $n_s$ 
and $\alpha_s$ at $k_0=0.05~{\rm Mpc}^{-1}$  also inflate. 

The geometric distortion effect helps to constrain $\Omega_{\rm m}$
from galaxy surveys alone. The radial distortion constrains
the expansion rate, $H(z)$, 
while the transverse distortion constrains 
the angular diameter distance, $D_A(z)$. 
Since $H(z)$ and $D_A(z)$ have different dependences on
 $\Omega_{\rm m}$ and $h$ for
a flat universe, the distortion can determine $\Omega_{\rm m}$ and $h $
 simultaneously ($\Omega_{\rm m}$ and $\Omega_{\rm m}h^2$ for our
 parameter set). 
The galaxy surveys at higher redshifts benefit
more from the geometric distortion effect. 

The redshift space distortion helps to constrain the neutrino
parameters, by lifting degeneracy between the neutrino parameters
and the galaxy bias.
We demonstrate it in the third block of Table~\ref{tab:noCMB}.
The errors on the neutrino parameters increase up to
nearly a factor of 5 for SG, if we ignore 
the information from the redshift space distortion.
The other parameters are not strongly affected.

\subsection{Variations with Survey Parameters}

\begin{figure}
  \begin{center}
    \leavevmode\epsfxsize=8.5cm \epsfbox{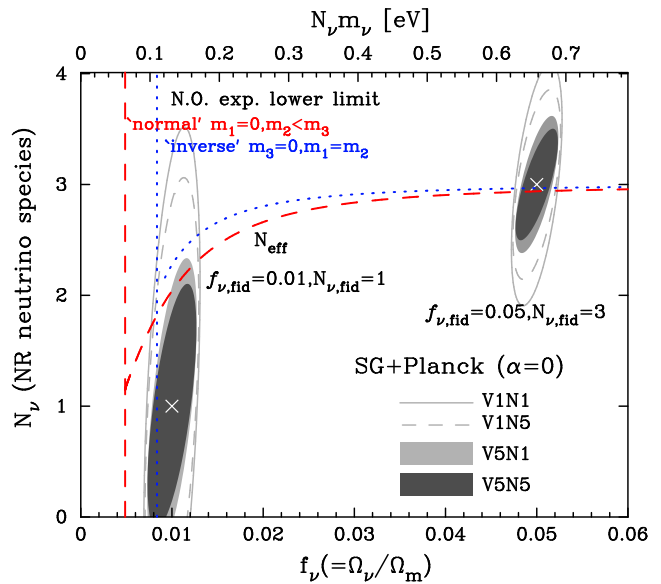}
  \end{center}
\vspace{-2em}
\caption{The same as Figure \ref{fig:fnu-Nnu}, but for advanced 
survey parameters for SG combined with Planck, as in Table \ref{tab:vn}.
``V1N1'' is the nominal survey parameters given in
 Table~\ref{tab:survey}.
 Note that zero running, $\alpha_s=0$, is assumed as a prior,
and  $f_{\nu,{\rm fid}}=0.05$ and $N^{\rm nr}_{\nu,{\rm fid}}=3$ are assumed.
}\label{fig:fnu-Nnu_VN}
\end{figure}

To guide the survey design, we show how parameter forecasts 
vary with two key survey parameters,
the number density of galaxies and the survey volume. 
Table \ref{tab:vn} shows how much 
one can reduce the projected errors on the key cosmological
parameters by increasing
the number density of galaxies or the survey volume or both by a factor of 5.
These advanced survey parameters are named as
\begin{itemize}
 \item V1N5: The survey volume is kept the same, while the number
       density of galaxies is
       increased by a factor of 5.
 \item V5N1: The survey volume is increased by a factor of 5, while the number
       density of galaxies is kept the same.
 \item V5N5: Both the survey volume and the number density
       of galaxies are increased by a factor of 5.
\end{itemize}
Figure
\ref{fig:fnu-Nnu_VN} shows how the error ellipses 
for $f_\nu$ and $N^{\rm nr}_\nu$ will shrink for these advanced
parameters.

We find that 
the most effective way to improve determination of the neutrino
parameters, particularly $N_\nu^{\rm nr}$, is to increase the survey volume.
One may understand this from Figure~\ref{fig:kmax} --- 
the information on the neutrino parameters saturates at 
$k_{\rm max}\sim 1~{\rm Mpc}^{-1}$, and thus there is not much to gain by reducing
the power spectrum errors at large $k$.

On the other hand, one can still improve determination of 
the parameters that determine the shape of the 
primordial power spectrum, $n_s$ and $\alpha_s$, by
increasing the number density of galaxies. This is especially true
for SG, which probes the largest $k_{\rm max}$.
Therefore, one can achieve even better accuracies for 
constraining inflationary models by either 
increasing the survey volume or the number density of galaxies, in principle;
however, in reality one is eventually going to be limited by our understanding
of the galaxy power spectrum in the weakly non-linear regime at large $k$.
In other words, there is not much to gain by reducing 
the power spectrum errors at large $k$, if the error is already
as small as theoretical uncertainty in the modeling of galaxy power 
spectrum at the same $k$.
Therefore, increasing the survey volume is probably the best way to
improve accuracies of both the neutrino parameters and the inflationary parameters.

\section{Conclusions and Discussions}
\label{conc}

The non-relativistic neutrinos and the tilt and running index of the
primordial power spectrum cause scale-dependent modifications in the
linear power spectrum probed by galaxy redshift surveys. We have
shown that one can determine these parameters  precisely by fully
exploiting the
two-dimensional information of galaxy clustering in angular and redshift
directions from high-redshift galaxy surveys, 
when combined with the CMB data from the Planck experiment.
The main results are summarized in Tables~\ref{tab:fnu005}
and ~\ref{tab:fnu001}, and may be graphically viewed in
Figures~\ref{fig:fnu-Nnu} and \ref{fig:alpha}.
Our conclusions are two fold.

The first conclusion is for the neutrino parameters.
We have found that the future galaxy surveys with 
$\Omega_{\rm s}=300$ deg$^2$ can provide very tight constraints on
the total neutrino mass.
The neutrino oscillation experiments have given the solid lower bound 
on the total neutrino mass, $m_{\nu, {\rm tot}}\simgt 0.06~$eV,
which is actually larger than the projected error on 
the total neutrino mass expected from the high-redshift galaxy surveys
we have considered, up to by a factor of 2.5 for 
the space-based survey targeting 
galaxies at $3.5<z<6.5$.
If two neutrino species have nearly equal masses 
(the inverted hierarchy), then the lower bound from the neutrino
oscillation experiments, $m_{\nu, {\rm tot}}\simgt 0.1~$eV, is up to 
4 times larger than the projected errors of cosmological experiments. 
Overall, 
the high-redshift galaxy surveys combined with Planck 
allow a positive {\em detection} of the total neutrino mass rather than the
upper limit, 
improving the constraints on 
the total neutrino mass by a factor of 20--40
compared with the current cosmological constraints.
The error on $m_{\nu, {\rm tot}}$ that we have found is smaller 
than that shown in the previous work (e.g., see \cite{HET}) 
by a factor of 5 -- 10, despite the fact that the survey
volume we assumed is larger than theirs only by a factor 1.5 -- 3. 
The main reason for the significant improvement is 
because our analysis exploits the full two-dimensional
information in the galaxy power spectrum in redshift space.
In particular,
the redshift space
distortion due to peculiar velocity field significantly helps 
improve the parameter determinations by breaking degeneracies 
between the cosmological parameters and the galaxy bias (also see Table
\ref{tab:noCMB}). 

In addition, we have carefully investigated how one can use the
future surveys to constrain the number of non-relativistic
neutrino species, which should
play an important role in resolving the neutrino mass hierarchy problem
as well as the neutrino absolute mass scale,
 independently of the total neutrino mass.
While we have found that the accuracy needed to discriminate between two
models with the same $m_{\nu, {\rm tot}}$ but different $N_\nu^{\rm nr}$,
$\sigma(N^{\rm nr}_\nu)\simlt 1$, is going to be difficult to 
achieve for the nominal survey designs we considered, 
one may achieve the desired precision
by enlarging the survey volume.
It should be stressed here that it is extremely important to exploit
independent constraints on $m_{\nu, {\rm tot}}$ and $N_\nu^{\rm nr} $ from
future cosmological observations, as any results unexpected from the
point of view of the standard three-flavor neutrino model would 
imply anomalies in our understanding of the neutrino physics and
hints for the new physics.
Needless to say, controlling the systematics in such observations
is also a crucial issue.

The second conclusion is for the shape of the primordial power spectrum.
We have graphically 
summarized the expected performance of the future galaxy surveys
for constraining the tilt, $n_s$, and the running index, $\alpha_s$, of the 
primordial power spectrum in Figure \ref{fig:alpha}. 
Compared with the constraints from the CMB data alone, 
the galaxy surveys we have considered can improve the determinations of 
$n_s$ and $\alpha_s$ by a factor of 2 and 3, 
yielding $\sigma(n_s)\sim 0.003$ and $\sigma(\alpha_s)\simeq 0.002$,
respectively.
The high-redshift galaxy surveys allow us to probe galaxy clustering in the
linear regime down to smaller length scales than at low redshifts. 
It is also important to note that the galaxy survey and CMB are
sensitive to the primordial power spectrum shape at different $k$
ranges, and therefore these two probes are nicely complementary to 
each other in terms of constraining the tilt and the running index 
at different $k$. We have explicitly shown that the
space-based galaxy survey (SG), such as CIP, has a similar level of
precision for constraining $n_s$ and $\alpha_s$ at the pivot scale $k_{\rm 0
pivot}\sim 0.5$ Mpc$^{-1}$, to the Planck experiment at $k_0=0.05~$Mpc$^{-1}$ 
(see Table \ref{tab:k0pivot}). 
While it is sometimes argued that simple inflationary models
should predict negligible amounts of running index which are out of 
reach of any experiments, our results show that the projected 
error on $\alpha_s$ from the SG survey
with five times more survey volume  is actually as small as 
$\alpha_s$ predicted by the simple model based on a massive,
self-interacting scalar field, $\alpha_s=-(0.8-1.2)\times 10^{-3}$.
Our results for $n_s$  and $\alpha_s$ are still on the conservative
side,  as we have ignored the baryonic oscillations in the analysis.
We have found that one can reduce the uncertainties in $n_s$ and $\alpha_s$
by a factor of 2--3 by including the baryonic oscillations --- however,
in that case one might also have to worry about the effect of 
curvature of the universe and dark energy properties.
We report the results of the full analysis elsewhere.

Throughout this paper, we have assumed the standard {\em three}-flavor
neutrinos, some of which have become non-relativistic by the present
epoch, as favored by neutrino oscillation experiments and the current
theory of particle physics. 
On the other hand,
cosmological observations such
as CMB and the large-scale structure can also put independent constraints
on the number of {\it relativistic}, weakly
interacting particles just like neutrinos, as a change in the
relativistic degrees of freedom directly affects the expansion rate
of the universe during the radiation era.
For example, an increase in the relativistic degrees of freedom 
delays the matter-radiation equality, to which the CMB and 
large-scale structure observables are sensitive.
The number of relativistic particles (minus photons) 
is conventionally expressed 
in terms of the effective number of neutrino species
(i.e., the temperature of additional relativistic species is assumed 
to be $(4/11)^{1/3}$ of the temperature of photons, just like ordinary
neutrinos).
The current cosmological bound is $N_{\nu}=4.2^{+1.2}_{-1.7}$ (95\%
C.L.) \cite{Hannestad05}, while the standard model predicts
$N_{\nu}=3.04$ \cite{QED} (also see \cite{Dodelson} 
for the constraint on
the abundance and mass of the sterile neutrinos based on the recent
cosmological data sets). 
Hence, it is interesting to explore how one can simultaneously constrain
the number of relativistic degrees of freedom before the photon
decoupling epoch  as well as the number of non-relativistic particle 
species at low redshifts, $N^{\rm nr}_\nu$, using
the future cosmological data sets.
Properties of such weakly interacting particles 
are still difficult to measure experimentally. 

It has been shown in the literature that 
the weak gravitational lensing \cite{TJ,Kev,Song} or the number 
count of galaxy clusters \cite{Fukugita,Haiman}
 can also be a powerful probe of the cosmological parameters. 
Different methods are sensitive to the structure formation at
different ranges of redshifts and wavenumbers, and have different
dependence on the cosmological parameters.
More importantly, they are subject to very different systematics.
Hence, by combining several  methods including the galaxy surveys
considered in this paper,
one can check for  systematics inherent in one particular method.
The combination of different methods may also reduce statistical
errors on the cosmological parameters.
While it is worth
exploring this issue carefully, we can easily imagine that, for the
neutrino parameters, the weak lensing or the cluster number count
does not improve $N_\nu^{\rm nr}$ very much.
To constrain $N_\nu^{\rm nr}$ better one has to find a way to
probe very large spatial scales, larger than the neutrino free-streaming
scales; however, these methods probe
only very small spatial scales where fluctuations have already become
non-linear. (They may provide improvements in the determination
of $m_{\nu,{\rm tot}}$.)

There is a promising way to check for systematics and improve 
the parameter determinations using
the galaxy survey alone.
While we have been assuming that perturbations are strictly linear
and hence perfectly Gaussian, in reality small non-linearity
always exists. The corrections to the power spectrum
due to such small non-linearity may be calculated analytically
using the higher-order cosmological perturbation
theory (see \cite{Bernardeau} for a review), 
which works extremely well at $z>2$
(Komatsu and Jeong, in preparation).
Using the same higher-order cosmological perturbation
theory, one may also compute the higher-order statistics,
such as the bispectrum, which is a very powerful tool to
check for systematics due to non-linearity in matter clustering,
redshift space distortion, and galaxy bias \cite{mat1,mat2}.
Also, 
the bispectrum and
power spectrum have different cosmological dependences \cite{TJ,dolney};
thus, it is naturally expected that combining the two would improve 
the determinations of cosmological parameters.  
The results of our investigation along these lines will be 
reported elsewhere.

No matter how powerful the bispectrum could be in terms of checking
for systematics, better models for non-linearity in redshift-space
distortion and galaxy bias are definitely required for our projected
errors to be actually realized.
A good news is that we do not need a fully non-linear description
of either component: we always restrict ourselves to the ``weakly
non-linear regime'' where perturbation theory should still be valid.
Having an accurate model for the redshift space distortion
in the weakly non-linear regime is important for the precise
determination of the total neutrino mass, as 
the redshift distortion plays a major role in lifting 
the degeneracy between the galaxy bias and the matter power
spectrum amplitude
(see Table~\ref{tab:noCMB} what happens when the redshift space
distortion is ignored).
Recently, a sophisticated model of the distortion effect
including the weak non-linear correction was
developed in \cite{Roman} based on the analytic method as well as the
simulations. Likewise, it will be quite possible to develop
a sufficiently accurate, well-calibrated model of the distortion effect
at least on large length scales, based on adequate simulations. 
While we have employed a scale-independent linear
bias throughout this paper, this model must break down even at
weakly non-linear regime.
Analytical \cite{mat3} as well as numerical \cite{white05,SE05,white05b}
work has shown that deviations from a
scale-independent bias do exist even on large spatial scales.
This effect would also become particularly important for the precision
measurements of the baryonic oscillations for constraining 
properties of dark energy.
Therefore, careful and systematic investigations based on
both analytical and numerical tools are needed to understand the 
realistic effect of scale-dependent bias on estimates of the cosmological
parameters.
As we have mentioned already, information from the higher-order
statistics in the galaxy clustering would be a powerful diagnosis
tool to check for systematics due to non-linear bias.
As perturbation theory predicts that the galaxy power spectrum 
and bispectrum should depend on the galaxy bias differently,  
one can directly determine the galaxy bias and cosmological parameters 
simultaneously, by combining the two statistical quantities 
\cite{mat1,mat2,dolney,Feldman,verde}. 
This method should also allow us to see a potential scale-dependent biasing
effect from the epoch of reionization \cite{babich/loeb:2005}.

Finally, let us comment on survey parameters. In order to make our
discussion general, we have considered three hypothetical surveys 
which are different in their redshift coverage and the number density
of targeted galaxies (see Sec.~\ref{survey} for the survey definition). 
We have also explored how the parameter errors would change when 
the survey parameters are varied from the fiducial values (see Table
\ref{tab:vn} and Figure \ref{fig:fnu-Nnu_VN}). Increasing a survey
volume has a greater impact on the parameter errors compared with
increasing the numbers of targeted galaxies for a given survey volume.
We are hoping that our results provide useful information to
help to define an optimal survey
design to attain the desired accuracy on the parameter determinations,
given the  limited observational resources and budget.

{\it Acknowledgments:} We thank K.~Ichiki, T.~Murayama, T.~Yoshida, 
and
M.~White for helpful discussions, and also thank an anonymous referee
 for useful
comments.
This work was supported in part by a
Grand-in-Aid for Scientific Research (17740129) of the Ministry of
Education, Culture, Sports, Science and Technology in Japan as well as
by the COE program at Tohoku University.  M.T. acknowledges the warm
hospitality  of University of Texas at Austin where this work was partly
done. 
E.K. acknowledges support from
the Alfred P. Sloan Foundation and NASA grant NNGO4GQ39G.  We
acknowledge the use of the publicly-available CMBFAST code
\cite{cmbfast}.

\appendix
\section{Properties of Cosmological Neutrinos}
\label{app:neutrino}
\subsection{Mass Density}

The present-day mass density of non-relativistic neutrinos 
is given by
\begin{equation}
 \rho^{\rm nr}_{\nu} = \sum_{i=1}^{N^{\rm nr}_\nu} m_{\nu,i}n_{\nu,i},
\end{equation}
where $i$ runs over the 
neutrino species 
{\it that are non-relativistic}, and $N_\nu^{\rm nr}$ is the number of 
non-relativistic neutrino species. 
We assume that some of the standard three active
neutrinos are massive and thus
 $0\leq N_\nu^{\rm nr}\leq 3$. 
These neutrinos were in thermal equilibrium with other particles
at early times until they decoupled from the primordial plasma slightly
before electron-positron annihilation.
Since they were still relativistic when they decoupled, 
their distribution function after decoupling is still given by 
that of a massless Fermion.
After electron-positron annihilation, the temperature of photons
became higher than the temperature of neutrinos by a factor of 
$(11/4)^{1/3}$.
Thus, the neutrino number density of each species, $n_{\nu}$, 
at a redshift relevant for a galaxy survey
 is given by the relativistic formula:
\begin{eqnarray}
 n_{\nu} &=& \frac{3\zeta(3)}{2\pi^2}T_\nu^3 \nonumber \\
&\simeq& 112(1+z)^3~{\rm cm}^{-3},
\end{eqnarray}
where 
$\zeta(3)\simeq 1.202$ and 
$T_\nu=(4/11)^{1/3}T_{\gamma 0}(1+z)$, where $T_{\gamma0}$ is the
present-day photon temperature and we have assumed $T_{\gamma0}=2.725$~K
\cite{mather}. 
Note that the number density  includes the contribution from
anti-neutrinos as well, and does not depend on neutrino species,
$n_\nu=n_{\nu,i}$.
The density parameter of non-relativistic neutrinos
is thus given by
\begin{eqnarray}
\nonumber
 \Omega_\nu &\equiv& \frac{8\pi G\rho^{\rm nr}_{\nu}}{3H_0^2}
=  \frac{8\pi G n_{\nu}}{3H_0^2}\sum_{i=1}^{N_\nu^{\rm nr}} m_{\nu,i}\\
&\simeq& \frac{\sum_i m_{\nu,i}}{94.1h^2~{\rm eV}}.
\label{eq:omnu}
\end{eqnarray}
Since $\Omega_\nu h^2$ must be less than the density parameter
of dark matter, $\Omega_{\rm cdm}h^2 \simeq 0.112$,  
the total mass of 
non-relativistic neutrinos must satisfy the following cosmological
bound,
\begin{equation}
 \sum_{i=1}^{N_\nu^{\rm nr}} m_{\nu,i} \lesssim 10.5~{\rm eV}.
\end{equation}

\subsection{Non-relativistic Epoch}

Neutrinos became non-relativistic when the mean energy
per particle given by
\begin{equation}
 \langle E\rangle = \frac{7\pi^4}{180\zeta(3)}T_\nu
\simeq 3.15T_\nu,
\end{equation}
fell below the mass energy, $m_{\nu,i}$.
The temperature at which a given neutrino species became
non-relativistic,
$T_{\nu,i}^{\rm nr}$, is thus given by
\begin{equation}
 T_{\nu,i}^{\rm nr} \equiv \frac{180\zeta(3)}{7\pi^4}m_{\nu,i}
\simeq 3680\left(\frac{m_{\nu,i}}{1~{\rm eV}}\right)~{\rm K}.
\end{equation}
The redshift at which a given neutrino species became non-relativistic
is
\begin{equation}
 1+z_{{\rm nr},i}
\simeq 1890\left(\frac{m_{\nu,i}}{1~{\rm eV}}\right).
\end{equation}
As the current constraints from the galaxy power spectrum at $z\sim 0$
already suggest $m_{\nu,i}\lesssim 1~{\rm eV}$, it is certain that
neutrinos became non-relativistic during the matter-dominated era.
The comoving wavenumber corresponding to the Hubble horizon size at
$z_{\rm nr}$
is thus given by
\begin{eqnarray}
\nonumber
 k_{{\rm nr},i} &\equiv& \frac{H(z_{{\rm nr},i})}{1+z_{{\rm nr},i}}
= \frac{\Omega^{1/2}_{m}h(1+z_{{\rm nr},i})^{1/2}}{2998~{\rm Mpc}}\\
&\simeq& 0.0145\left(\frac{m_{\nu,i}}{1~{\rm
		eV}}\right)^{1/2}\Omega^{1/2}_{m}h~{\rm Mpc}^{-1}.
\label{eq:knr}
\end{eqnarray}
Note that this value is smaller than that given in \cite{HET}
by a factor of $\simeq \sqrt{3.15}$. They assumed that 
neutrinos became non-relativistic when
$T_\nu=m_\nu$, rather than $T_\nu=(180\zeta(3)/7\pi^2)m_\nu\simeq m_\nu/3.15$.

\subsection{Neutrino Free-streaming Scale}

Density perturbations of non-relativistic neutrinos grow only 
when the comoving wavenumber of perturbations is below the 
free-streaming scale, $k_{\rm fs}$, given by
\begin{equation}
 k_{{\rm fs},i}(z) \equiv \sqrt{\frac32}\frac{H(z)}{(1+z)\sigma_{v,i}(z)},
\label{eq:jeans}
\end{equation}
where $\sigma^2_{v,i}(z)$ is the velocity dispersion of neutrinos and given
in \cite{BES} as
\begin{eqnarray}
\nonumber
 \sigma^2_{v,i}(z) &\equiv& 
\frac{\int \frac{d^3p~p^2/m^2}{\exp[p/T_{\nu}(z)]+1}}{\int \frac{d^3p}{\exp[p/T_{\nu}(z)]+1}}
= \frac{15\zeta(5)}{\zeta(3)}\frac{T_\nu^2(z)}{m_{\nu,i}^2}\\
&=& \frac{15\zeta(5)}{\zeta(3)}\left(\frac{4}{11}\right)^{2/3}\frac{T_\gamma^2(0)(1+z)^2}{m_{\nu,i}^2},
\end{eqnarray}
where $\zeta(5)\simeq 1.037$. 
Hence,
\begin{equation}
 k_{{\rm fs},i}(z) \simeq \frac{0.677}{(1+z)^{1/2}}
\left(\frac{m_{\nu,i}}{1~{\rm eV}}\right)
\Omega_{\rm m}^{1/2}h~{\rm Mpc}^{-1}.
\label{eq:kfs}
\end{equation}
Neutrino density perturbations with $k>k_{{\rm fs},i}$ cannot grow because
pressure gradient prevents neutrinos from collapsing gravitationally;
thus, neutrinos are effectively smooth at $k>k_{{\rm fs},i}$,
and the power spectrum of neutrino perturbations is exponentially suppressed.
Note that Eq.~(\ref{eq:jeans}) is exactly the same as 
the Jeans scale in an expanding universe for collisional particles,
if $\sigma_v$ is replaced by the speed of sound.

\section{Inflationary Predictions}
\label{app:inflation}

\subsection{Generic Results}

Inflationary predictions are commonly expressed in terms of 
the shape of the primordial power spectrum of curvature 
perturbations in comoving gauge, ${{\cal R}}$:
\begin{equation}
 \langle {{\cal R}}_{\mathbf k}{{\cal R}}^*_{{\mathbf k}'}\rangle
= (2\pi)^3
P_{\cal R}(k)
\delta^{(3)}({\mathbf k}-{\mathbf k}'),
\end{equation}
where
\begin{equation}
\frac{k^3P_{\cal R}(k)}{2\pi^2}
=\delta_{\cal R}^2\left(\frac{k}{k_0}
\right)^{-1+n_s+\frac12\alpha_s\ln(k/k_0)}.
\label{eqn:delta_zeta}
\end{equation}
Here, $n_s$ and $\alpha_s$ are called the ``tilt'' and ``running index''
of the primordial power spectrum. These parameters are related to the
shape of potential, $V(\phi)$, of an {\it inflaton} field, $\phi$, a field which caused
inflation, as \cite{liddle/lyth:2000}
\begin{eqnarray}
\label{eq:ns}
 n_s -1 &=& M_{\rm pl}^2\left[- 3\left(\frac{V'}{V}\right)^2 
+ 2\left(\frac{V''}{V}\right)\right],\\
\nonumber
 \alpha_s &=& 2M_{\rm pl}^4\left[4\left(\frac{V'}{V}\right)^2\left(\frac{V''}{V}\right)
-3\left(\frac{V'}{V}\right)^4\right. \\
& & \left.- \left(\frac{V'V'''}{V^2}\right)\right],
\label{eq:alphas}
\end{eqnarray}
where primes denote derivatives with respect to $\phi$,
and $M_{\rm pl}\equiv (8\pi G)^{-1/2}=2.4\times 10^{18}$~GeV is the reduced
Planck mass. Successful inflationary models must yield sufficiently large
number of $e$-folds for the expansion of the universe, $N$, before
inflation ends at $t_{\rm end}$ when $\phi$ rolls down on the potential to
$\phi=\phi_{\rm end}$:
\begin{equation}
 N = \int_t^{t_{\rm end}}dt~H(t) \approx 
\frac1{M_{\rm pl}^2}\int_{\phi_{\rm end}}^\phi d\phi\frac{V}{V'},
\end{equation}
which has to be at least as large as 50. This condition  
requires $|n_s-1|$ and $|\alpha_s|$ to
be much less than unity, while exact values depend on specific
inflationary models (i.e., the shape of $V(\phi)$). 
Therefore, precision determination of $n_s$ and $\alpha_s$ is a very
powerful tool for constraining inflationary models.

\subsection{Specific Examples}

An illustrative example of inflationary models is inflation
caused by a massive, self-interacting real scalar field:
\begin{equation}
 V(\phi) = \frac12m_\phi^2\phi^2 + \frac14\lambda\phi^4,
\end{equation}
where $m_\phi$ is the mass of $\phi$ and $\lambda$ ($>0$) is the coupling
constant
of self-interaction. 
The mass term and the interaction term equal when $\phi=\phi_c$, where
\begin{equation}
 \phi_c\equiv m_\phi\sqrt{\frac{2}{\lambda}}.
\end{equation}
The mass term dominates when $\phi< \phi_c$, while the interaction term
dominates when $\phi> \phi_c$.
One obtains
\begin{eqnarray}
\label{eq:Nex}
 N &\approx& 
\frac{\phi^2}{8M_{\rm pl}^2}
+ \frac{\phi_c^2}{16M_{\rm pl}^2}\ln\left(1+2\frac{\phi^2}{\phi_c^2}\right),\\
\nonumber
 n_s -1 &=& 
-\frac{8M_{\rm pl}^2}{\phi^2}\\
\label{eq:nsex}
&\times  &
\frac{1+\frac52\frac{\phi^2}{\phi_c^2}+3\frac{\phi^4}{\phi_c^4}}{\left(1+\frac{\phi^2}{\phi_c^2}\right)^2},\\
\nonumber
 \alpha_s &=& -\frac{32M_{\rm pl}^4}{\phi^4}\\
&\times &
\frac{\left(1+2\frac{\phi^2}{\phi_c^2}\right)
\left(1+3\frac{\phi^2}{\phi_c^2}+2\frac{\phi^4}{\phi_c^4}+3\frac{\phi^6}{\phi_c^6}\right)}
{\left(1+\frac{\phi^2}{\phi_c^2}\right)^4}.
\label{eq:alphasex}
\end{eqnarray}

Let us take the limit of mass-term driven inflation, $\phi\ll \phi_c$.
One finds
\begin{eqnarray}
 N &\rightarrow& \frac{\phi^2}{4M_{\rm pl}^2},\\
 n_s -1 &\rightarrow& -\frac{8M_{\rm pl}^2}{\phi^2}
= -\frac{2}{N},\\
 \alpha_s &\rightarrow& -\frac{32M_{\rm pl}^4}{\phi^4}
= -\frac{2}{N^2}.
\end{eqnarray}
For $N=50$, $n_s=0.96$ and $\alpha_s=-0.8\times 10^{-3}$.
In the opposite limit, self-coupling driven inflation, $\phi\gg \phi_c$,
one finds
\begin{eqnarray}
 N &\rightarrow& \frac{\phi^2}{8M_{\rm pl}^2},\\
 n_s -1 &\rightarrow& -\frac{24M_{\rm pl}^2}{\phi^2}
= -\frac{3}{N},\\
 \alpha_s &\rightarrow& -\frac{192M_{\rm pl}^4}{\phi^4}
= -\frac{3}{N^2}.
\end{eqnarray}
For $N=50$, $n_s=0.94$ and $\alpha_s=-1.2\times 10^{-3}$.
These simple examples show that a precision of 
$\sigma(n_s)\sim 10^{-3}$ is sufficient
for discriminating between models, while $\sigma(\alpha_s)\sim 10^{-3}$
may allow us to detect $\alpha_s$ from simple (though not the simplest) 
inflationary models driven
by a massive scalar field with self-coupling.

\section{Normalizing Power Spectrum}
\label{app:norm}

While inflation predicts the power spectrum of ${{\cal R}}$,
what we observe from galaxy redshift surveys is the power spectrum
of matter density fluctuations, $\delta_m$. In this section we derive the
conversion from ${{\cal R}}$ during inflation to $\delta_m$ at a
particular redshift after the matter-radiation equality.

Let us begin by writing the Poisson equation in Fourier space,
\begin{equation}
 -k^2 \Psi_{\mathbf k}(a) = 4\pi G\rho_m(a) \delta_{m,{\mathbf k}}(a)
  a^2
=  \frac{3H_0^2\Omega_{\rm m}}{2} \frac{\delta_{m,{\mathbf k}}(a)}{a}.
\end{equation}
where $\Psi$ is gravitational potential in the usual (Newtonian) sense
($g_{00}=-1+2\Psi$). 
Cosmological perturbation theory
relates $\Psi$ after the matter-radiation equality 
to ${{\cal R}}$ during inflation as 
\begin{equation}
 \Psi_{\mathbf k}(a) = -\frac35{{\cal R} }_{\mathbf k}T(k)\frac{D_{cb\nu}(k,a)}{a},
\end{equation}
where 
$D_{cb\nu}(k,a)$ is the linear growth factor of total matter
perturbations including CDM, baryons and non-relativistic neutrinos,
and $T(k)$ is the linear transfer function.
Note that the transfer function and the growth rate
are normalized such that $T(k)\rightarrow 1$ as $k\rightarrow 0$
and $D_{cb\nu}/a\rightarrow 1$ as $k\rightarrow 0$ during the matter era.
(We assume that neutrinos became non-relativistic during the
matter-dominated era.)
Hence,
\begin{equation}
  \delta_{m,{\mathbf k}}(a)
= \frac{2k^2}{5H^2_0\Omega_{\rm m}}{{\cal R} }_{\mathbf k}T(k){D_{cb\nu}(k,a)},
\end{equation}
which gives the power spectrum of $\delta_m$,
\begin{equation}
 P(k,a)
= \left(\frac{2k^2}{5H^2_0\Omega_{\rm m}}\right)^2
P_{{\cal R}}(k)T^2(k)D^2_{cb\nu}(k,a),
\label{eq:pknorm1}
\end{equation}
after the matter-radiation equality.
The WMAP collaboration has determined the normalization of 
$k^3P_{\cal R}(k)/(2\pi^2)$ at $k=k_0=0.05$~Mpc$^{-1}$ as
\begin{equation}
 \delta_{\cal R}^2 = 2.95\times 10^{-9}A,
\end{equation}
where $A$ is a constant of order unity
(Eq.~[32] of \cite{verde/etal:2003}). 
Putting these results together,
we finally obtain the power spectrum of density perturbations normalized
to WMAP:
\begin{eqnarray}
\nonumber
 \frac{k^3P(k,z)}{2\pi^2}
&=& 2.95\times10^{-9}A \left(\frac{2k^2}{5H^2_0\Omega_{\rm m}}\right)^2\\
&& \times
{D^2_{cb\nu}(k,z)}
T^2(k)\left(\frac{k}{k_0}\right)^{-1+n_s+\frac12\alpha_s\ln(k/k_0)}.
\label{eq:pknorm}
\end{eqnarray}

\section{Correlations between parameter estimates}
\label{correl}

It is worth noting how the parameter estimations are correlated with
each other for a given survey, which can be quantified by the
correlation coefficients defined by Eq.~(\ref{eqn:coeff}).  Table
\ref{tab:coeff} gives the correlation matrix for the parameters for SG
combined with Planck.
\begin{table*}
\begin{center}
\begin{tabular}{l|cccccccccccc}
\hline\hline
& $\Omega_{\rm m}$ & $\delta_{\cal R}$
& $f_\nu$ & $N_{\nu}^{\rm nr}$ & $n_s$ &
$\alpha_{\rm s}$ & $\Omega_{\rm m}h^2$
& $\Omega_{\rm b}h^2$ & $\tau$ 
& $b_1(z=4)$& $b_1(z=5)$& $b_1(z=6)$
\\ \hline
$\Omega_{\rm m}$ & 1 & 0.079 & -0.28 & -0.79 & -0.68 & -0.32 & 0.94 &
 -0.41 & -0.073 & -0.96 & -0.96 & -0.96\\
$\delta_{\cal R}$ & 0.079& 1  & 0.81 & 0.24 & 0.10 & -0.17 & 0.090 & -0.001 &
 0.98  & -0.14 & -0.16 & -0.17\\
$f_\nu$& -0.28& 0.81& 1 & 0.39 & 0.55 & -0.26 & -0.27 & 0.033 & 0.86 & 0.30 &
 0.28 & 0.27\\
$N_\nu^{\rm nr}$ & -0.79& 0.24&0.39 & 1 & 0.30 & 0.45 & -0.70& 0.43 & 0.34 & 
 0.79 & 0.78 & 0.78 \\ 
$n_{\rm s}$ & -0.68& 0.10&0.55 & 0.30& 1 & -0.29 & -0.64 & 0.25 & 0.24 & 0.64 & 0.64 & 0.63\\ 
$\alpha_{\rm s}$ & -0.32& -0.17& -0.26& 0.45& -0.29& 1 & -0.29 & 0.17 & -0.13 & 0.35 & 0.35 & 0.35\\ 
$\Omega_{\rm m}h^2$ & 0.94& 0.090& -0.27& -0.70& -0.64& -0.29& 1 & -0.10 & -0.073 & -0.90 & -0.90 & -0.90\\ 
$\Omega_{\rm b}h^2$ & -0.41& -0.001& 0.033& 0.43& 0.25& 0.17& -0.10& 1 & 0.022 & 0.40 & 0.40 & 0.40\\ 
$\tau$ & -0.073& 0.98& 0.86& 0.34& 0.24&-0.13 & -0.073&0.022 & 1 & 0.0071 & -0.0086 & -0.021\\ 
$b_1(z=4)$ & -0.96& -0.14& 0.30& 0.79& 0.64& 0.35& -0.90& 0.40& 0.0071& 1 & 0.99 & 0.99\\ 
$b_1(z=5)$ & -0.96& -0.16& 0.28& 0.78& 0.64& 0.35& -0.90& 0.40& -0.0086& 0.99& 1 & 0.99\\ 
$b_1(z=6)$ & -0.96& -0.17& 0.27& 0.78& 0.63& 0.35& -0.90& 0.40& -0.021& 0.99& 0.99& 1 \\ 
\hline\hline
\end{tabular}
\end{center}
\caption{The correlation matrix for parameter estimation errors for SG
 combined with Planck. }
\label{tab:coeff}
\end{table*}



\begin{thebibliography}{99}
%
%
\bibitem{WMAP}
  C.~Bennet et al., \apj~ Suppl. {\bf 148}, 1 (2003).
\bibitem{Tegmark}
  M.~Tegmark et al., \prd~ {\bf 69}, 103501 (2004).
\bibitem{Seljak}
  U.~Seljak et al., \prd~ {\bf 71}, 103515 (2005). 
\bibitem{jarvis/jain:2005}
  M.~Jarvis, B.~Jain, G.~Bernstein and D.~Dolney, astro-ph/0502243.
\bibitem{Peebles}
  P.~J.~E.~Peebles and J.~T.~Yu, \apj~ {\bf 162}, 815 (1970).
\bibitem{SZ}
  R.~A.~Sunyaev and Ya.~B.~Zel'dovich, Astrophys. Sp. Sci.~ {\bf 7}, 3 (1970).
\bibitem{BE}
  J.~R.~Bond and G.~Efstathiou, \apj~Lett.~ {\bf 285}, 45 (1984).
\bibitem{liddle/lyth:2000}
  A.~R.~Liddle and D.~H.~Lyth, {\em Cosmological Inflation and the
	Large-Scale Structure}, Cambridge University Press (2000).
\bibitem{dodelson}
  S.~Dodelson,  {\em Modern Cosmology} Academic Press, San Diego, (2003).
\bibitem{WMAPcl}
  G.~Hinshaw et al., \apj~ Suppl. {\bf 148}, 135 (2003).
\bibitem{WMAPte}
  A.~Kogut et al., \apj~ Suppl. {\bf 148}, 161 (2003).
\bibitem{bao1}
  D.~J.~Eisenstein et al., \apj~ {\bf 633}, 560 (2005).
\bibitem{bao2}
  S.~Cole, \MNRAS~ {\bf 362}, 505 (2005).
\bibitem{sn1}
  A.~G.~Riess et al., Astron. J.~ {\bf 116}, 1009 (1998).
\bibitem{sn2}
  S.~Perlmutter et al., \apj~ {\bf 517}, 565 (1999).
\bibitem{DEreview}
  P.~J.~E.~Peebles and B.~Ratra, Rev.~Mod.~Phys.~{\bf 75}, 559 (2003).
\bibitem{NGreview}
  N.~Bartolo, E.~Komatsu, S.~Matarrese and A.~Riotto, 
  Phys.~Rept.~ {\bf 402}, 103 (2004).
\bibitem{2dF}
  W.~J.~Percival et al., \MNRAS~ {\bf 327}, 1297 (2001).
\bibitem{Peiris}
  H.~V.~Peiris, et al., \apj~ Suppl. {\bf 146}, 213 (2003). 
\bibitem{exp1}
  S.~ Fukuda et al. [Super-Kamiokande Collaboration], \prl~ {\bf 85}, 3999 (2000).
\bibitem{exp2}
  S.~N.~Ahmed et al. [SNO Collaboration], \prl~ {\bf 92}, 181301 (2004).
\bibitem{exp3}
  K.~Eguchi et al. [KamLAND Collaboration], \prl~ {\bf 90}, 021802 (2003).
\bibitem{exp4} 
  T.~Arakki et al. [KamLAND Collaboration], \prl~ {\bf 94} 081801 (2005).
\bibitem{exp5}
  R.~D.~McKeown and P.~Vogel, Phys.~Rep.~ {\bf 394}, 315 (2004). 
\bibitem{BES}
  J.~R. Bond, G.~Efstathiou and J.~Silk, \prl~ {\bf 45}, 1980 (1980).
\bibitem{BS}
  J.~R.~Bond and A.~Szalay, \apj~ {\bf 276}, 443 (1983).
\bibitem{Pog1}
  D.~Pogosyan and A.~Starobinsky, \MNRAS~ {\bf 265}, 507 (1993)
\bibitem{Pog2}
  D.~Pogosyan and A.~Starobinsky, \apj~ {\bf 447}, 465 (1995)
\bibitem{Ma95}
  C.-P.~Ma and E.~Bertschinger, \apj~ {\bf 455}, 7 (1995).
\bibitem{Ma}
  C.-P.~Ma, \apj~ {\bf 471}, 13 (1996).
\bibitem{HET}
  W.~Hu, D.~Eisensttein and M.~Tegmark, Phys.~Rev.~Lett. {\bf 80}, 5255 (1998).
\bibitem{Hannestad1}
  S.~Hannestad, \prd~ {\bf 67}, 085017 (2003).
\bibitem{Lahav}
  \O.~Elgar\o y and O.~Lahav, New.J.Phys. {\bf 7}, 61 (2005).
\bibitem{LahavSuto}
  O.~Lahav and Y.~Suto, Living Reviews in Relativity~ {\bf 7}, 1
	(2004). 
\bibitem{Hannestad2}
  S.~Hannestad, astro-ph/0511595.
\bibitem{WMAPp}
  D.~N.~Spergel et al., \apj~ Suppl. {\bf 148}, 175 (2003).
\bibitem{Bonn}
  J.~Bonn et al., Nucl.Phys.B (Proc. Suppl.) {\bf 91}, 273 (2001).
\bibitem{LSND}
  A.~Aguilar et al. [LSND Collaboration], \prd~ {\bf 64}, 112007 (2001).
\bibitem{SE}
  H.-J.~Seo and D.~Eisenstein, \apj~ {\bf 598}, 720 (2003).
\bibitem{Taka03}
  T.~Matsubara and A.~Szalay, \prl~ {\bf 90}, 021302 (2003)
\bibitem{HuHaiman}
  W.~Hu and Z.~Haiman, \prd~ {\bf 68}, 063004 (2003). 
\bibitem{Blake}
  C.~Blake and K.~Glazebrook, \apj~ {\bf 594}, 665 (2003). 
\bibitem{GB05}
  K.~Glazebrook	and C.~Blake, \apj~ in press (astro-ph/0505608).
\bibitem{taka05}
  T.~Matsubara, \apj~ {\bf 615}, 573 (2004). 
\bibitem{Yamamoto}
  K.~Yamamoto, B.~A.~Bassett and H.~Nishioka, \prl~ {\bf 94}, 051301 (2005). 
\bibitem{fmos}
  M.~Kimura et al.,  
  Proceedings of the SPIE, {\bf 4841}, 974 (2003)
\bibitem{wfmos}
  K.~Glazebrook et al., astro-ph/0507457. 
\bibitem{hetdex}
  G.~J.~Hill, K.~Gebhardt, E.~Komatsu and P.~J.~MacQueen,
  AIP Conf. Proc., {\bf 743}, 224 (2004).
\bibitem{cip}
  G.~J.~Melnick et al., 
  The NASA Origins Probe Mission Study Report, 
  ``{\it The Cosmic Inflation Probe: Study Report}'' (2005).
  See also {\sf http://cfa-www.harvard.edu/cip}.
%
%
\bibitem{mather}
  J.~C.~Mather et al., \apj~ {\bf 512}, 511 (1999).
\bibitem{HE98}
  W.~Hu and D.~Eisensttein, \apj~ {\bf 498}, 497 (1998).
\bibitem{HE99} 
  W.~Hu and D.~Eisenstein, \apj~ {\bf 511}, 5 (1999).
\bibitem{TJ}
  M.~Takada and B.~Jain, \MNRAS~ {\bf 348}, 897 (2004).
\bibitem{AP}
  C.~Alcock and B.~Paczynski, \nat, {\bf 281} 358 (1979). 
\bibitem{Taka96}
  T.~Matsubara and Y.~Suto, \apj{}Lett., {\bf 470}, 1 (1996).
\bibitem{Ballinger}
  W.~E.~Ballinger, J.~A.~Peacock and A.~F.~Heavens, \MNRAS, {\bf 282}, 877
	(1996).
\bibitem{Kaiser}
  N.~Kaiser, \MNRAS~ {\bf 227}, 1 (1987).
\bibitem{FKP}
  H.~A.~Feldman, N.~Kaiser and J.~A.~Peacock, \apj~ {\bf 426}, 23 (1994).
\bibitem{Tegmark97}
  M.~Tegmark, A.~N.~Taylor and A.~F.~Heavens, \apj~ {\bf 480}, 22 (1997). 
\bibitem{Meiksin}
  A.~Meiksin, M.~White and J.~A.~Peacock, \MNRAS~ 304, 851 (1999).
\bibitem{white05}
  M. White, Astropart.~Phys.~ {\bf 24}, 334 (2005).
\bibitem{SE05}
  H.-J.~Seo and D.~Eisenstein, \apj~ {\bf 633}, 575 (2005).
%
%
\bibitem{Hu02}
  W.~Hu, \prd~ {\bf 65}, 023003 (2002).
\bibitem{HJ04}
  W.~Hu and B.~Jain, \prd~ {\bf 70}, 043009 (2004). 
%
%
\bibitem{cmbfast}
  U.~Seljak and M.~Zaldarriaga, \apj~ {\bf 469}, 437 (1996).
\bibitem{EHT}
D.~Eisenstein, W.~Hu and M.~Tegmark, \apj~ {\bf 518}, 2 (1998).
\bibitem{Ouchi}
  M.~Ouchi, K.~Shimasaku, S.~Okamura et al., \apj~ {\bf 611} 660 (2004).
%
%
\bibitem{Hannestad05}
  S.~Hannestad,  astro-ph/0510582.
\bibitem{Dodelson}
  S.~Dodelson, A.~Melchiorri and A.~Slosar, astro-ph/0511500. 
\bibitem{QED}
  A.~D.~Dolgov, Phys.~Rep.~{\bf 370}, 333 (2002). 
\bibitem{Roman}
  R.~Scoccimarro, \prd~ {\bf 70}, 083007 (2004)
\bibitem{Kev}
  K.~Abazajian and S.~Dodelson, \prl~ {\bf 91}, 041301 (2003).
\bibitem{Song}
  Y.~S.~Song and L.~Knox, \prd~ {\bf 70}, 063510 (2004).
\bibitem{Fukugita}
  M.~Fukugita, G.-C. Liu and N. Sugiyama, \prl~ {\bf 84}, 1082 (2000).
\bibitem{Haiman}
  S.~Wang, Z.~Haiman, W.~Hu, J.~Khoury and M.~Morgan, \prl~ {\bf 95}
  011302 (2005).
\bibitem{Bernardeau}
  F.~Bernardeau, S.~Colombi, E.~Gaztanaga and R.~Scoccimarro, 
  Phys.~Rep.~{\bf 367}, 1 (2002). 
\bibitem{mat1}
  S.~Matarrese, L.~Verde and A.~F.~Heavens, \MNRAS~ {\bf 290}, 651 (1997).
\bibitem{mat2}
  L.~Verde, A.~F.~Heavens and S.~Matarrese, \MNRAS~ {\bf 300}, 747 (1998).
\bibitem{dolney}
  D.~Dolney, B.~Jain and M.~Takada, \MNRAS~ in press (astro-ph/0409455).
\bibitem{mat3}
  A.~F.~Heavens, L.~Verde,  and S.~Matarrese, \MNRAS~ {\bf 301}, 797 (1998).
\bibitem{white05b}
  A.~E.~Schulz and M.~White, astro-ph/0510100. 
\bibitem{Feldman}
  H.~A.~Feldmann, J.~A.~Frieman, J.~N.~Fry and R.~Scoccimarro, \prl~{\bf 86}, 1434 (2001).
\bibitem{verde} 
  L.~Verde et al., \MNRAS~{\bf 335}, 432 (2002). 
\bibitem{babich/loeb:2005}
  D.~Babich and A.~Loeb, astro-ph/0509784.
%
%
\bibitem{verde/etal:2003}
  L.~Verde, et al., \apj~ Suppl.~ {\bf 148}, 195 (2003). 
\end{thebibliography}
\end{document}